\documentclass[12pt,letterpaper]{article}
\pdfoutput=1
\usepackage{ifpdf}
\usepackage[a4paper,top=3.2cm,width=17cm, bottom=3.2cm, left=2cm]{geometry}

\usepackage{graphicx}
\usepackage{microtype}
\usepackage[utf8x]{inputenc}
\usepackage[T1]{fontenc}
\usepackage{ae}
\usepackage{aecompl}
\usepackage{ytableau}
\usepackage{hyperref}
\usepackage{amsthm}
\usepackage{amsmath,calc,mathtools}
\usepackage{amssymb}
\usepackage{fullpage}
\usepackage{amsmath}
\usepackage{cleveref}
\usepackage{comment}
\usepackage{authblk}
\usepackage{tikz-cd}
\usepackage{quiver}
\usepackage{arydshln}
\usepackage{fdsymbol}
\usepackage{faktor}
\usepackage{adjustbox}
\usepackage{mathrsfs}
\usepackage{appendix}
\usepackage{hyperref}
\hypersetup{
    colorlinks=true,
    linkcolor=blue,
    filecolor=magenta,      
    urlcolor=cyan,
    citecolor=blue
}
\usepackage{authblk} 

\definecolor{amber}{rgb}{1.0, 0.49, 0.0}
\definecolor{Green}{rgb}{0.0, 0.5, 0.0}
\definecolor{purple}{rgb}{0.7,0,0.7}

\makeatletter
\newcommand*\dashline{\rotatebox[origin=c]{90}{$\dabar@\dabar@\dabar@$}}
\makeatother

\DeclareMathOperator{\diag}{diag}
\DeclareMathOperator{\coeff}{coeff}
\DeclareMathOperator{\lowest}{lowest\_degree}

\newtheorem{thm}{Theorem}

\newtheorem{cor}[thm]{Corollary}
\newtheorem{lma}[thm]{Lemma}

\newtheorem{dfn}[thm]{Definition}

\title{Quiver diagonalization and open BPS states}

\author[1]{Jakub Jankowski\thanks{jakub.jankowski@uwr.edu.pl}}

\author[2]{Piotr Kucharski\thanks{piotr.kucharski@mimuw.edu.pl}}

\author[3,4]{H\'{e}lder Larragu\'{i}vel\thanks{helder.larraguivel@uj.edu.pl}}

\author[5]{\break Dmitry Noshchenko\thanks{d.noshchenko@uva.nl}}

\author[4]{Piotr Su{\l}kowski\thanks{psulkows@fuw.edu.pl}}

\affil[1]{Institute of Theoretical Physics, University of Wroc{\l}aw, pl. Borna 9, 50-204 Wroc{\l}aw, Poland}
\affil[2]{Institute of Mathematics, University of Warsaw, ul. Banacha 2, 02-097 Warsaw, Poland}
\affil[3]{Institute of Physics, Jagiellonian University, ul. {\L}ojasiewicza 11, 30-348 Krak\'{o}w, Poland}
\affil[4]{Faculty of Physics, University of Warsaw, ul. Pasteura 5, 02-093 Warsaw, Poland}
\affil[5]{Institute of Physics, University of Amsterdam, Science Park 904, 1098 XH, Amsterdam, the~Netherlands}

\begin{document}
\thispagestyle{empty}
\maketitle
\begin{abstract}
   We show that motivic Donaldson-Thomas invariants of a~symmetric quiver~$Q$, captured by the~generating function $P_Q$, can be encoded in another quiver $Q^{(\infty)}$ of (almost always) infinite size, whose only arrows are loops, and whose generating function $P_{Q^{(\infty)}}$ is equal to $P_Q$ upon appropriate identification of generating parameters. Consequences of this statement include a~generalization of the~proof of integrality of Donaldson-Thomas and Labastida-Mari\~{n}o-Ooguri-Vafa invariants that count open BPS states, as well as expressing motivic Donaldson-Thomas invariants of an~arbitrary symmetric quiver in terms of invariants of $m$-loop quivers. In particular, this means that the~already known combinatorial interpretation of invariants of $m$-loop quivers extends to arbitrary symmetric quivers. 
\end{abstract}

\newpage 

\tableofcontents

\newpage

\section{Introduction and summary}

Quivers play a~pivotal role in mathematics and physics. For mathematicians, an~important challenge is to understand the~structure of moduli spaces of quiver representations. This structure is characterized by various invariants, in particular motivic Donaldson-Thomas (DT) invariants. In physics, one role of quivers is to characterize BPS states in supersymmetric theories -- in this context, nodes of a~quiver correspond to certain basic states, while arrows encode how these basic states may form more complicated bound states. Such states are enumerated by (motivic) DT invariants (or appropriate combinations thereof) and their multiplicities provide an~important information about a~given theory. 

In this paper we focus on symmetric quivers, meaning that for each arrow connecting two different nodes, there is also an~arrow in the~opposite direction. Symmetric quivers are understood better than generic quivers, and in particular it is known how to determine their (motivic) DT invariants \cite{KS0811,KS1006,Efi12,Rei12,FR1512,MR1411}. Symmetric quivers are especially relevant in the~knots-quivers correspondence \cite{KRSS1707short,KRSS1707long}, where they are assigned to knots and encode BPS spectra in associated 3-dimensional $\mathcal{N}=2$ theories. All this provides an~important motivation for this work.

The main result of this paper is the~statement that motivic DT invariants of an~arbitrary symmetric quiver can be expressed in terms of motivic DT invariants of another quiver, generically of infinite size, whose only arrows are loops (loops are arrows that connect a~node to itself).
If we encode a~structure of a~quiver in a~matrix $C$, whose entry $C_{ij}$ is the~number of arrows from node $i$ to $j$, then the~matrix representing the~latter (generically infinite) quiver is diagonal -- this is why we introduce the~name diagonal quiver. 

\begin{dfn}
    We call the~quiver diagonal, if all of its arrows are loops, i.e. all nodes are disconnected. 
\end{dfn}

Recall that motivic DT invariants of a~symmetric quiver $Q$ are encoded in the~factorization of the~quiver generating series $P_Q(\boldsymbol{x},q)$, whose detailed form is given in (\ref{PQ}), and which depends on a~quiver matrix $C$, a~number of generating parameters $(x_1,\dots,x_{|Q_0|})=\boldsymbol{x}$ with each $x_i$ associated to the~$i$'th node, and the~motivic parameter $q$. The~main result of this paper, i.e. a~relation between motivic DT invariants of a~symmetric quiver and the~corresponding diagonal quiver, follows in fact from the~relation between their generating functions, which is given in the~following theorem. (Note that matrices $C$ that we consider may also have negative entries; to formalize this feature, we introduce extended quivers, which are objects corresponding to such matrices.)

\begin{thm}
\label{thrm:main}
For every symmetric extended quiver $Q$ there exist a~diagonal extended quiver $Q^{(\infty)}$ and a~set of identifications $\{ x^{(\infty)}_i = x^{(\infty)}_i(x_1,\dots,x_{|Q_0|},q) \}_{i \in Q^{(\infty)}_0}$ 
such that the~motivic generating series of $Q$ and $Q^{(\infty)}$ are equal upon the~identification of generating parameters given by this set:
    \begin{equation}
        P_{Q}(\boldsymbol{x},q)=\left.P_{Q^{(\infty)}}(\boldsymbol{x^{(\infty)}},q)\right|_{x^{(\infty)}_i = x^{(\infty)}_i(x_1,\dots,x_{|Q_0|},q)}.
    \end{equation}
\end{thm}

This theorem has further interesting consequences. Note that motivic DT invariants of a~diagonal quiver are expressed in terms of motivic DT invariants of $m$-loop quivers, i.e. quivers which consist of one node and $m$ loops. Therefore, our main result relates motivic DT invariants of an~arbitrary symmetric quiver to those of $m$-loop quivers. Since invariants of the $m$-loop quiver, as well as its Cohomological Hall Algebra, are quite well understood \cite{Rei12}, it is of a~great advantage to connect them to invariants of an~arbitrary symmetric quiver. In particular, it is quite interesting to know the~interpretation of the~factorization of the~motivic generating series of an~arbitrary quiver into generating series for $m$-loop quivers at the~level of Cohomological Hall Algebras. On the~other hand, taking advantage of this factorization and the~knowledge of classical DT invariants for $m$-loop quivers, it would be of interest to derive expressions for classical DT invariants for an~arbitrary symmetric quiver that were proposed in \cite{PSS1802}. Similarly, a~combinatorial interpretation of invariants of $m$-loop quivers extends now to an~arbitrary symmetric quiver. Relation to $m$-loop quivers also provides a~novel proof of integrality of motivic DT invariants of a~symmetric quiver. This relation has also an~interesting physical interpretation in the~context of 3-dimensional $\mathcal{N}=2$ theories. We discuss some of these aspects in this paper and leave the~rest for elucidation in future work.

The plan of the~paper is as follows. In Sec. \ref{sec-quiversDT} we summarize relevant results for symmetric quivers and generalize the~description of $m$-loop case (with $m\in \mathbb{N})$ to negative values $-m$. In Sec. \ref{sec:Diagonalization} we present how to determine a~diagonal quiver corresponding to an~arbitrary symmetric quiver. In Sec. \ref{sec:DT} we discuss how to express (motivic) DT invariants of a~symmetric quiver in terms of invariants of the~corresponding diagonal quiver. In Sec.~\ref{sec:simple} , \ref{sec:KQ correspondence} and \ref{sec:FK quivers} we illustrate our results respectively in simple examples, examples related to the~knots-quivers correspondence, and for the~variant of this correspondence related to $F_K$ invariants. In App. \ref{app:combinatorial} we discuss combinatorial properties of $(-m)$-loop quivers.

\section{Symmetric quivers and DT invariants}  \label{sec-quiversDT}

In this section we recall relevant aspects of the~study of symmetric quivers and (motivic) DT invariants. We also explain the~generalization of quiver arrows that allows for the~negative entries in the~adjacency matrix and show that DT invariants of an~$(-m)$-loop quiver are in one-to-one correspondence with those of the~$(m+1)$-loop quiver.

\subsection{Motivic generating series}

Quiver $Q$ is a~pair $(Q_0,Q_1)$, where $Q_0$ is a~set of vertices and $Q_1$ is a~set of arrows $i\rightarrow j$. We denote the~number of arrows between vertices $i$ and $j$ by $C_{ij}$, and assemble these numbers into $|Q_0|\times |Q_0|$ adjancency matrix $C$. A~quiver is called symmetric if for each arrow between two different vertices there is also an~arrow in the~opposite direction; this also means that $C_{ij}=C_{ji}$.

For a~given quiver, it is important to understand the~structure of moduli spaces of its representations, i.e. spaces of linear maps $\mathbb{C}^{d_i}\to\mathbb{C}^{d_j}$, where each $\mathbb{C}^{d_i}$ is assigned to vertex~$i$. $\boldsymbol{d}=(d_1,\ldots,d_{|Q_0|})\in \mathbb{N}^{|Q_0|}$ is called the~dimension vector. Basic information about such spaces is encoded in their Betti numbers or their generalizations, which are captured by motivic DT invariants $\Omega_{(d_1,\ldots,d_{|Q_0|}),s}=\Omega_{\boldsymbol{d},s}$. For a~symmetric quiver, these invariants are encoded in the~motivic generating series
\begin{equation}
    P_Q(\boldsymbol{x},q) =
    \sum_{\boldsymbol{d}}(-q)^{\boldsymbol{d} \cdot C\cdot\boldsymbol{d}}\frac{\boldsymbol{x}^{\boldsymbol{d}}}{(q^{2};q^{2})_{\boldsymbol{d}}}
    = \sum_{d_1,\dots,d_{|Q_0|}\geq 0}(-q)^{\sum_{i,j=1}^{|Q_0|} C_{ij}d_i d_j}\prod_{i=1}^{|Q_0|}\frac{x_i^{d_i}}{(q^{2};q^{2})_{d_i}},    \label{PQ}
\end{equation}
where a~generating parameter $x_i$ is assigned to each vertex $i\in Q_0$ and 
\begin{equation}
    (\alpha;q^{2})_{n} = \prod_{k=0}^{n-1}(1-\alpha q^{2k})
\end{equation}
is the~$q$-Pochhammer symbol. The~product decomposition of this series into quantum dilogarithms determines $\Omega_{\boldsymbol{d},s}$ as follows\footnote{Note that in this paper the~sign is included in the~definition of $\Omega$'s and the~product form with quantum dilogarithms serves a~basis. In consequence, we have $\sum_{s\in\mathbb{Z}}\Omega_{\boldsymbol{d},s}q^s=\sum_{j\in\mathbb{Z}}\Omega'_{\boldsymbol{d},j}(-q)^{j+1}$, where $\Omega'_{\boldsymbol{d},j}$ is the~DT invariant in the~notation from \cite{KRSS1707long}.}
\begin{equation}\label{PQ-product}
    P_Q(\boldsymbol{x},q) = \prod_{\boldsymbol{d},s}(\boldsymbol{x}^{\boldsymbol{d}}q^s;q^2)_{\infty}^{\Omega_{\boldsymbol{d},s}} = \prod_{\boldsymbol{d}\in \mathbb{N}^{|Q_0|}\setminus \boldsymbol{0}} \prod_{s\in\mathbb{Z}} \prod_{k\geq 0} \Big(1 - (x_1^{d_1}\cdots x_{|Q_0|}^{d_{|Q_0|}}) q^{2k+s} \Big)^{\Omega_{(d_1,\ldots,d_{|Q_0|}),s}}.
\end{equation}
We also introduce a~generating series
\begin{equation}
    \Omega(\boldsymbol{x},q) = \sum_{\boldsymbol{d}} \Omega_{\boldsymbol{d}}(q)\, \boldsymbol{x}^{\boldsymbol{d}} =  \sum_{\boldsymbol{d}\in \mathbb{N}^{|Q_0|}\setminus \boldsymbol{0}}\sum_{s\in \mathbb{Z}} \,\Omega_{(d_1,\ldots,d_{|Q_0|}),s}\, x_1^{d_1}\cdots x_{|Q_0|}^{d_{|Q_0|}} q^s.   \label{Omega-series}
\end{equation}
%
%
A non-trivial fact is that invariants $\Omega_{(d_1,\ldots,d_{|Q_0|}),s}$ defined via the~decomposition (\ref{PQ-product}) are integer, and multiplied by $(-1)^{j+1}$ become positive \cite{KS1006,Efi12}. 


In fact, in this work we consider a~larger class of quiver matrices $C$, which may also have negative entries. Note that such quiver matrices appear in the~knots-quivers correspondence \cite{KRSS1707long}. To take such cases into account, we slightly generalize the~usual definition of a~quiver.

\begin{dfn}\label{def:extended quiver}
Let $C
$ be an~arbitrary integer symmetric matrix. We say that it defines an~extended quiver $Q=(Q_0,Q_1^{+}\cup Q_1^{-})$ where $Q_0$ is a~set of nodes, $Q_1^{+}$ is a~set of arrows and $Q_1^{-}$ is a~set of negative arrows, such that
\begin{itemize}
    \item every $C_{ii}>0$ ($C_{ii}<0$) corresponds to $|C_{ii}|$ loops (negative loops) at the~node $i$, 
    \item every $C_{ij}>0$ ($C_{ij}<0$) corresponds to $|C_{ij}|$ pairs of arrows
    (negative arrows) between $i$ and $j$ in $Q$.
\end{itemize}
\end{dfn}

As an~example, consider extended quiver
\begin{equation}
    Q = 
    \begin{tikzcd}
    	\bullet & \bullet & \bullet
    	\arrow[curve={height=-6pt}, from=1-1, to=1-2]
    	\arrow[curve={height=-6pt}, from=1-2, to=1-1]
    	\arrow[curve={height=-6pt}, dashed, from=1-2, to=1-3]
    	\arrow[curve={height=-6pt}, dashed, from=1-3, to=1-2]
    \end{tikzcd}
    \,,
    \qquad
    C =
    \left[
    \begin{array}{ccc}
         0 & 1 & 0 \\
         1 & 0 & -1 \\
         0 & -1 & 0
    \end{array}
    \right]
    \,.
\end{equation}
The pair of ordinary arrows is denoted by solid lines, whereas the~pair of negative arrows is denoted by dashed lines. Moreover, if all $C_{ij}\geq 0$, we call $Q$ a~proper quiver (or simply quiver), whereas if all $C_{ij}\leq 0$, we call $Q$ a~negative quiver.

\subsection{Unlinking and linking}



Basing on the~study presented in \cite{EKL1910}, we know that
for any symmetric quiver $Q$ with $|Q_0|$ nodes,
there exists another symmetric quiver with $|Q_0|+1$ nodes, given by a~removal (or an~addition)
of a~pair of arrows in $Q$ and addition of an~extra node\footnote{Essentially, this procedure is based on the~enumerative-geometric interpretation of motivic DT generating series as a~generating series for open Gromov-Witten invariants, as noted in \cite{EKL1811} and \cite{EKL1910}}. Importantly,
 motivic DT invariants for the~two quivers are equal after a~proper identification of a~variable for the~new node. We call such operations unlinking and linking, and
define them as follows.

\begin{dfn}[Ekholm, Kucharski, Longhi]\label{def:(un)linking}
Consider a~symmetric extended quiver $Q$ together with a~set of generating parameters $\{ x_i \}_{i\in Q_{0}}$ and fix $a,b\in Q_{0}$. 
\begin{itemize}
    \item The~unlinking of nodes $a,b$ is defined as a~transformation of $Q$ leading to a~new quiver $Q^{(\text{unlinked})}$ and a~set of identifications $\{x^{(\text{unlinked})}_i=x^{(\text{unlinked})}_i(x_1,\dots,x_{|Q_0|},q)\}$ such that:
\begin{itemize}
\item There is a~new node $n$: $Q^{(\text{unlinked})}_{0}=Q_{0}\cup \{n\}$.

\item The~number of arrows of the~new quiver is given by
\begin{align}
\label{eq:unlinking arrows}
C^{(\text{unlinked})}_{ab} & =C_{ab}-1, & C^{(\text{unlinked})}_{nn} & =C_{aa}+2C_{ab}+C_{bb}-1,\\
C^{(\text{unlinked})}_{in} & =C_{ai}+C_{bi}-\delta_{ai}-\delta_{bi}, & C^{(\text{unlinked})}_{ij} & =C_{ij}\quad\textrm{for all other cases,} \nonumber
\end{align}
where $\delta_{ij}$ is a~Kronecker delta.
\item The~functions encoding the~identification of generating parameters are given by
\begin{equation}
\begin{split}
\label{eq:unlinking change of vars}
x^{(\text{unlinked})}_{n}(x_1,\dots,x_{|Q_0|},q)&=q^{-1}x_{a}x_{b},\\ x^{(\text{unlinked})}_{i}(x_1,\dots,x_{|Q_0|},q)&=x_i\qquad \forall i\neq n.   
\end{split} 
\end{equation}
\end{itemize}
\item The~linking of nodes $a,b$ is defined as a~transformation of $Q$ leading to a~new extended quiver $Q^{(\text{linked})}$ such that:
\begin{itemize}
\item There is a~new node $n$: $Q^{(\text{linked})}_{0}=Q_{0}\cup \{n\}$.
\item The~number of arrows of the~new quiver is given by
\begin{align}
\label{eq:nlinking arrows}
C^{(\text{linked})}_{ab} & =C_{ab}+1, & C^{(\text{linked})}_{nn} & =C_{aa}+2C_{ab}+C_{bb},\\
C^{(\text{linked})}_{in} & =C_{ai}+C_{bi}, & C^{(\text{linked})}_{ij} & =C_{ij}\quad\textrm{for all other cases~.} \nonumber
\end{align}
\item The~functions encoding the~identification of generating parameters are given by
\begin{equation}
\begin{split}
x^{(\text{linked})}_{n}(x_1,\dots,x_{|Q_0|},q)&=x_{a}x_{b},\\ x^{(\text{linked})}_{i}(x_1,\dots,x_{|Q_0|},q)&=x_i\qquad \forall i\neq n.   
\end{split} 
\end{equation}
\end{itemize}
\end{itemize}
\end{dfn}

\begin{thm}[Ekholm, Kucharski, Longhi]\label{thm:(un)linking}
The~operations of unlinking and linking
both preserve the~motivic generating function of the~quiver:
\begin{equation}
\begin{split}
    P_{Q}(\boldsymbol{x},q)
    &=\left.P_{Q^{(\text{unlinked})}}(\boldsymbol{x}^{(\text{unlinked})},q)\right|_{x^{(\text{unlinked})}_i = x^{(\text{unlinked})}_i(x_1,\dots,x_{|Q_0|},q)},\\
    &=\left.P_{Q^{(\text{linked})}}(\boldsymbol{x}^{(\text{linked})},q)\right|_{x^{(\text{linked})}_i = x^{(\text{linked})}_i(x_1,\dots,x_{|Q_0|},q)}.
\end{split}
\end{equation}

\end{thm}

For brevity, in the~rest of the~paper we will write (un)linking whenever we mean either unlinking or linking.

\subsection{\texorpdfstring{$m$}{m}-loop quivers}\label{subsec:m-loop}

In this section we summarize properties of $m$-loop quivers, i.e. quivers that consist of one vertex and $m\in \mathbb{N}$ loops.  
In what follows we generalize some properties of $m$-loop quivers to extended quivers (which may be thought of as possibly having also negative number of loops, denoted by $-m$), and show that DT invariants of a~$(-m)$-loop quiver are in one-to-one correspondence with those of $(m+1)$-loop quiver.

An adjacency matrix of an~$m$-loop quiver consists of a~single entry, which counts the~number of loops:
%
\begin{equation}
\begin{tabular}{ccc}
    \vspace{1cm} \\
    $Q = \qquad\quad$
    &
    \begin{tikzpicture}[scale=2, every loop/.style={}, baseline=(current  bounding                           box.center),overlay]
        \draw [fill] (0, 0) circle [radius=0.02];
        \node [draw=none] {} edge [in=50-20,out=130+20,loop,scale=5] ();
        \node [draw=none] {} edge [in=50-10,out=130+10,loop,scale=3.5] ();
        \node [draw=none] {} edge [in=50+20,out=130-20,loop,scale=2] ();
        \node at (0.27, 0.5) { $\cdots$ };
    \end{tikzpicture}
    & $\qquad, \qquad\quad C=[\,m \,]\,$.
\end{tabular}
\end{equation}
The motivic generating series of such a~quiver takes form
\begin{equation}\label{eq:m-loop}
P_{\text{$m$-loop}}(x,q) = 
\sum_{d=0}^{\infty} \frac{(-q)^{md^2}}{(q^2;q^2)_d}\,x^d\,.
\end{equation}
From the~perspective of this paper, an~important feature of $m$-loop quivers is that they play a~role of building blocks of diagonal quivers. 

The case $m\geq 0$ has been studied in relation to moduli of quiver representations \cite{Rei12,Rei11}, as well as counting of topological strings and quantum knot invariants \cite{KRSS1707short, KRSS1707long}. A~product decomposition (\ref{PQ-product}) for the~series (\ref{eq:m-loop}), which encodes motivic Donaldson-Thomas (DT) invariants $\Omega^m_{r,s}\in \mathbb{Z}$ \cite{KS0811,KS1006,Efi12}, in this case takes form
\begin{equation}\label{eq:m-loop-factorized}
    P_{m-\rm loop}(x,q) = \prod_{r,s}\left(x^r q^s;q^2\right)^{\Omega^m_{r,s}}_{\infty}\, .
\end{equation}
We stress again that in our convention $\Omega^m_{r,s}$ differ from the~usual DT invariants considered in the~literature by absorbing a~minus sign -- we do so for future convenience.
The generating series of DT invariants (\ref{Omega-series}) in this case takes form
\begin{equation}
    \Omega_{\text{$m$-loop}}(x,q) = \sum_{r=1}^{\infty} \Omega^{m}_r(q)\, x^r =  \sum_{r,s} \Omega^m_{r,s}\, x^r q^s \, . 
\end{equation}
The product (\ref{eq:m-loop-factorized}) is finite only when $m=0$ and $m=1$:

\begin{equation}
    P_{\text{0-loop}}(x,q) = \frac{1}{(x;q^2)_{\infty}}, \quad P_{\text{1-loop}}(x,q) = 
    (qx;q^2)_{\infty}\,
\end{equation}
which corresponds to
\begin{equation}
    \Omega_{\text{0-loop}}(x,q) = -x,\quad \Omega_{\text{1-loop}}(x,q) = qx\, .
\end{equation}
Otherwise, the~combinatorial structure encoded in DT invariants is quite involved and the~spectrum of DT invariants is infinite.

In \cite{Rei12} a~combinatorial model which captures the~behavior of DT invariants for $m$-loop quivers with $m\geq 0$ was introduced. Equivalent model, which is especially convenient for us, has been described in \cite{KS1608} -- we discuss it in more detail in App. \ref{app:combinatorial}. 

We write the~generating function of DT invariants for a~few $m$-loop quivers in Tab.~\ref{tab:mqPoch}.
\begin{table}[h!]
    \centering
    \renewcommand{\arraystretch}{1.5}
    \begin{tabular}{||c|c||}
    \hline
$m$ &\ $\Omega_{\text{$m$-loop}}(x,q)$ \\
\hline
\hline
0 &\ $-x$ \\
\hline
1 &\ $qx$ \\
\hline
2 &\ $-q^2x + q^4x^2 - q^8x^3 + O(x^4)$ \\
\hline
3 &\ $q^3x + q^8x^2 + (q^{11}+q^{13}+q^{17})\, x^3 + O(x^4)$ \\
\hline
4 &\ $-q^4x + (q^8+q^{12})x^2 - (q^{14}+q^{16}+q^{18}+q^{20}+q^{22}+q^{26})\, x^3
+ O(x^4) $ \\
\hline
5 &\ $q^5x + (q^{12}+q^{16})x^2 + (q^{17}+q^{19}+q^{21}+2q^{23}+q^{25}+q^{27}+q^{29}+$ \\
&\ $q^{31}+q^{35})\, x^3 + O(x^4) $ \\ 
\hline
\end{tabular}
    \caption{Generating functions of DT invariants for $m$-loop quivers with $m\geq 0$.}
    \label{tab:mqPoch}
\end{table}

\noindent
One can notice that 
\begin{equation}
\label{eq:m-loop DT}
    \Omega_{\text{$m$-loop}}(x,q) = (-1)^{m-1}q^mx + O(x^2),
\end{equation}
which is true for any $m\in\mathbb{N}$ (in fact, as we will see in Lemma \ref{lma:mloop}, it is true also for any $m\in\mathbb{Z}$).

Now our goal is to extend (\ref{eq:m-loop}) to negative integers $-m$. This is important, since an~infinite diagonal quiver can have negative entries.
We also note that it would be very interesting to understand a~possible representation-theoretic meaning of this generalization.
We proceed with an~important lemma generalizing (\ref{eq:m-loop-factorized}) to $(-m)$-loop quivers.

\begin{lma}
\label{lma:mloop}
For any $m\in\mathbb{N}$, the~$(-m)$-loop quiver generating series admits the~following form
\begin{equation}\label{eq:negative-loop-factorized}
    P_{\text{\rm $(-m)$-loop}}(x,q) = \prod_{r,s}\left(x^r q^s;q^2\right)^{\Omega^{-m}_{r,s}}_{\infty}~,
\end{equation}
where
\begin{equation}
    \Omega^{-m}_r(q) = -q^{2-r}\, \Omega^{m+1}_r(q^{-1})~.
    \label{eq:BPS_iso}
\end{equation}
\end{lma}

We provide two independent proofs of the~above lemma. The~one presented below is rather elementary and involves a~simple $q$-Pochhammer identity. The~second one is included in App. \ref{app:combinatorial} and it shows how this structure is captured by the~combinatorial models from \cite{Rei12,KS1608}. 

\begin{proof}

We use the~following $q$-Pochhammer identity

\begin{equation}\label{eq:q-identity}
    (\alpha;q^2)_d = (-1)^d q^{d^2-d} \alpha^d (\alpha^{-1};q^{-2})_d\, ,
\end{equation}
where $\alpha$ is any formal variable, to obtain
\begin{equation}\label{eq:cminusid}
    P_{\text{$(-m)$-loop}}(x,q^{-1}) = \sum_{d\geq 0}\frac{(-q)^{md^2}}{(q^{-2};q^{-2})_d}x^d = 
    \sum_{d\geq 0} \frac{ (-q)^{(m+1)\,d^2}}{(q^2;q^{2})_d}(qx)^d = P_{\text{$(m+1)$-loop}}(qx,q)\, .
\end{equation}
Applying this result to $m=0$ gives
\begin{equation}\label{eq:qPoch-identity}
   (x;q^{-2})_{\infty} = \frac{1}{(q^2x;q^2)_{\infty}}\, .
\end{equation}

The latter in turn can be applied to every $q$-Pochhammer in (\ref{eq:m-loop-factorized}), yielding the~explicit relation between the~DT invariants (\ref{eq:BPS_iso}).
\end{proof}

We finish the~discussion of $(-m)$-loop quivers by tabulating DT invariants for some values of $-m$ (Tab. \ref{tab:mqPoch-negative}). A~complete combinatorial description of these invariants is presented in App. \ref{app:combinatorial}.

\begin{table}[h!]
    \centering
    \renewcommand{\arraystretch}{1.5}
    \begin{tabular}{||c|c||}
    \hline
$-m$ &\ $\Omega_{\text{$(-m)$-loop}}(x,q)$ \\
\hline
\hline
$-1$ &\ $q^{-1}x + q^{-4}x^2 + q^{-9}x^3 + O(x^4)$ \\
\hline
$-2$ &\ $-q^{-2}x - q^{-8}x^2 - (q^{-18} + q^{-14} + q^{-12})\, x^3 + O(x^4)$ \\
\hline
$-3$ &\ $q^{-3}x - (q^{-12} + q^{-8})\, x^2 + (q^{-27} +  q^{-23} + q^{-21} + q^{-19} + q^{-17} + q^{-15})\, x^3 + O(x^4)$ \\
\hline
$-4$ &\ $ -q^{-4}x - (q^{-16} + q^{-12})\, x^2 - (q^{-36}+q^{-32}+q^{-30}+q^{-28}+q^{-26}+2q^{-24}+q^{-22}+ $ \\
&\ $q^{-20}+q^{-18})\, x^3 + O(x^4)$ \\
\hline
\end{tabular}
    \caption{Generating functions of DT invariants for $(-m)$-loop quivers.}
    \label{tab:mqPoch-negative}
\end{table}

\section{Quiver diagonalization}
\label{sec:Diagonalization}

In this section we describe what we call a~\emph{quiver diagonalization}. We start from an~extended symmetric quiver $Q$ and successively apply (un)linking operations, ending up with a~(generically infinite) diagonal extended quiver, denoted $Q^{(\infty)}$. The~only arrows of $Q^{(\infty)}$ are loops, and thus its motivic generating series factorizes
as a~product of $(\pm m)$-loop quivers. We also discuss the~uniqueness of this infinite quiver.

\subsection{\texorpdfstring{$n$-th}{nth} degree approximation}
\label{sec:nth degree approximation}

We will construct $Q^{(\infty)}$ order by order, so we start from the~following:

\begin{dfn}
    We say that a~pair consisting of a~symmetric extended quiver $Q^{(n)}$ and a~set of identifications $\{x^{(n)}_i=x^{(n)}_i(x_1,\dots,x_{|Q_0|},q)\}_{i\in Q^{(n)}_0}$ is an~$n$-th degree approximation of a~symmetric extended quiver~$Q$, if  
    \begin{equation}
    P_{Q}(\boldsymbol{x},q)-  \left.P_{Q^{(n)}}(\boldsymbol{x}^{(n)},q)\right|_{x^{(n)}_i=x^{(n)}_i(x_1,\dots,x_{|Q_0|},q)}
    = O(\boldsymbol{x}^{n+1}),
    \end{equation}
    i.e. the~difference contains only terms proportional to $\boldsymbol{x}^{\boldsymbol{d}}=\prod_{i=1}^{|Q_0|}x_i^{d_i}$ with total degree $|\boldsymbol{d}|=\sum_{i}d_{i}\geq n+1$. 
\end{dfn}

\begin{thm}\label{thm:Qn-approx}
For every symmetric extended quiver $Q$ and $n\in\mathbb{Z}_{+}$ there exists an~$n$-th degree approximation of $Q$ such that the~extended quiver $Q^{(n)}$ is diagonal and finite.
\end{thm}

\begin{proof}
Let us fix a~symmetric quiver $Q$. Looking at the~general expression
for the~motivic generating series
\begin{equation}\label{eq:Quiver series}
P_{Q}(\boldsymbol{x},q)=\sum_{\boldsymbol{d}}(-q)^{\boldsymbol{d} \cdot C\cdot\boldsymbol{d}}\frac{\boldsymbol{x}^{\boldsymbol{d}}}{(q^{2};q^{2})_{\boldsymbol{d}}},
\end{equation}
we can see that the~lowest degree contribution of each diagonal entry
of the~quiver adjacency matrix to the~quiver motivic generating series
is
\begin{equation}
(-q)^{C_{ii}}\frac{x_{i}}{(q^{2};q^{2})_{1}},
\end{equation}
which is of total degree 1. On the~other hand, the~lowest degree contribution
of each non-diagonal entry to $P_{Q}$ is
\begin{equation}
(-q)^{2C_{ij}}\frac{x_{i}x_{j}}{(q^{2};q^{2})_{1}^{2}}~,
\end{equation}
which is of total degree 2. From \cite{EKL1910}, we know that the~lowest
degree contribution of the~new diagonal entry $C_{new.new}$ that
comes from the~unlinking of $C_{ij}$ is 
\begin{equation}
(-q)^{C_{new.new}}\frac{x_{new}}{(q^{2};q^{2})_{1}}=(-q)^{C_{ii}+C_{jj}+2C_{ij}-1}\frac{q^{-1}x_{i}x_{j}}{(q^{2};q^{2})_{1}},
\end{equation}
whereas the~one that comes from linking of $C_{ij}$ is 
\begin{equation}
(-q)^{C_{new.new}}\frac{x_{new}}{(q^{2};q^{2})_{1}}=(-q)^{C_{ii}+C_{jj}+2C_{ij}}\frac{x_{i}x_{j}}{(q^{2};q^{2})_{1}}~.
\end{equation}
In both cases it is of total degree 2, since we count the~degree in the~variables $x_{i}$ corresponding to the~\emph{initial} quiver $Q$.
We can generalize these considerations and write that each non-diagonal
entry $C_{rs}$ and the~new entry coming from the~unlinking or linking of nodes $r$ and $s$ both have the~lowest degree contribution equal to the~sum of the~lowest degree contributions of nodes $r$ and $s$. In
consequence, the~lowest degree of the~contribution to $P_{Q}$ -- a~positive integer -- can be assigned to each entry of the~adjacency matrix $C$ and matrices coming from the~unlinking or linking procedure.

Now we will recursively construct $\tilde{C}^{(n)}$ and $C^{(n)}$
-- the~adjacency matrices of quivers $\tilde{Q}^{(n)}$ and $Q^{(n)}$
-- assigning to each entry of the~matrix the~lowest degree of the
contribution to $P_{Q}$. We start from assigning $1$ to each diagonal
entry and 2 to each non-diagonal entry of $C$. In order to avoid
confusion with matrix entries, we will write assigned lowest degrees
in brackets:
\begin{equation}
\lowest(C)=\left[\begin{array}{ccccc}
(1) & (2) & (2) & \cdots & (2)\\
(2) & (1) & (2) & \ddots & \vdots\\
(2) & (2) & (1) & \ddots & (2)\\
\vdots & \ddots & \ddots & \ddots & (2)\\
(2) & \cdots & (2) & (2) & (1)
\end{array}\right].
\end{equation}
Then, we unlink all positive non-diagonal entries of $C$ and link all negative non-diagonal entries. We denote the~resulting matrix $\tilde{C}^{(1)}$ and the~corresponding change of variables~$\tilde x^{(1)}_i=\tilde x^{(1)}_i(x_1,\dots,x_{|Q_0|},q)$
\begin{equation}\label{eq:LowestDegree(1)}
\lowest(\tilde{C}^{(1)})=\left[\begin{array}{ccccccc}
(1) &  &  & (3) & (3) & \cdots & (3)\\
 & \ddots &  & \vdots & \vdots & \ddots & \vdots\\
 &  & (1) & (3) & (3) & \cdots & (3)\\
(3) & \cdots & (3) & (2) & (4) & \cdots & (4)\\
(3) & \cdots & (3) & (4) & (2) & \ddots & \vdots\\
\vdots &  & \vdots & \vdots & \ddots & \ddots & (4)\\
(3) & \cdots & (3) & (4) & \cdots & (4) & (2)
\end{array}\right].
\end{equation}
Moreover, we can see that the~$|Q_0|\times |Q_0|$ top-left corner of matrix
$\tilde{C}^{(1)}$ is a~diagonal
matrix which we will call $C^{(1)}$ (and the~corresponding quiver
$Q^{(1)}$). Since it came from unlinking or linking all non-diagonal entries of the~matrix $C$, we have
\begin{equation}
\label{eq:1st order approx}
C^{(1)}=\left[\begin{array}{cccc}
\tilde{C}_{11}^{(1)} & 0 & \cdots & 0\\
0 & \tilde{C}_{22}^{(1)} & \ddots & \vdots\\
\vdots & \ddots & \ddots & 0\\
0 & \cdots & 0 & \tilde{C}_{|Q_0||Q_0|}^{(1)}
\end{array}\right]=\left[\begin{array}{cccc}
C_{11} & 0 & \cdots & 0\\
0 & C_{22} & \ddots & \vdots\\
\vdots & \ddots & \ddots & 0\\
0 & \cdots & 0 & C_{|Q_0||Q_0|}
\end{array}\right]
, \quad
x^{(1)}_i=x_i
\end{equation}
for all $i\in Q_0$, and
\begin{equation}
\label{eq:PQ for 1st order approx}
\begin{split}
P_{Q}(\boldsymbol{x},q)&=\left.P_{\tilde{Q}^{(1)}}(\tilde{\boldsymbol{x}}^{(1)},q)\right|_{\tilde x^{(1)}_i=\tilde x^{(1)}_i(x_1,\dots,x_{|Q_0|},q)}\\
&=1+\sum_{i=1}^{|Q_0|}(-q)^{C_{ii}}\frac{x_{i}}{(q^{2};q^{2})_{1}}+O(\boldsymbol{x}^{2})=\left.P_{Q^{(1)}}(\boldsymbol{x}^{(1)},q)\right|_{x^{(1)}_i=x_i}+O(\boldsymbol{x}^{2}),
\end{split}
\end{equation}
so $(Q^{(1)},\{x^{(1)}_i=x_i\}_{i\in {Q}_0})$ is a~first degree approximation of $Q$.

Now we move to the~induction step and assume that for some $n\in\mathbb{Z_{+}}$ there exists an~$n$-th degree approximation of $Q$ formed by a~finite diagonal quiver $Q^{(n)}$ and a~change of variables $\{x^{(n)}_i=x^{(n)}_i(x_1,\dots,x_{|Q_0|},q)\}_{i\in Q^{(n)}_0}$, as well as a~finite symmetric quiver $\tilde{Q}^{(n)}$ obtained from the~(un)linking of $Q$, which contains $Q^{(n)}$ as a~subquiver\footnote{The change of variables $\{\tilde x^{(n)}_i=\tilde x^{(n)}_i(x_1,\dots,x_{|Q_0|},q)\}_{i\in Q^{(n)}_0}$ follows from the~changes of variables associated to (un)linking, given in Def. \ref{def:(un)linking}.}. In other words
\begin{equation}
\begin{split}
C^{(n)}&=\left[\begin{array}{cccc}
\tilde{C}_{11}^{(n)} & 0 & \cdots & 0\\
0 & \tilde{C}_{22}^{(n)} & \ddots & \vdots\\
\vdots & \ddots & \ddots & 0\\
0 & \cdots & 0 & \tilde{C}_{kk}^{(n)}
\end{array}\right],
\\
\tilde{C}^{(n)}&=\left[\begin{array}{cccccc}
 &  &  & \tilde{C}_{1,k+1}^{(n)} & \cdots & \tilde{C}_{1,|\tilde{Q}^{(n)}_0|}^{(n)}\\
 & C^{(n)} &  & \vdots &  & \vdots\\
 &  &  & \tilde{C}_{k,k+1}^{(n)} & \cdots & \tilde{C}_{k,|\tilde{Q}^{(n)}_0|}^{(n)}\\
\tilde{C}_{1,k+1}^{(n)} & \cdots & \tilde{C}_{k,k+1}^{(n)} & \tilde{C}_{k+1,k+1}^{(n)} & \cdots & \tilde{C}_{k+1,|\tilde{Q}^{(n)}_0|}^{(n)}\\
\vdots &  & \vdots & \ddots & \ddots & \vdots\\
\tilde{C}_{1\tilde{k}}^{(n)} & \cdots & \tilde{C}_{k,|\tilde{Q}^{(n)}_0|}^{(n)} & \tilde{C}_{k+1,|\tilde{Q}^{(n)}_0|}^{(n)} & \cdots & \tilde{C}_{|\tilde{Q}^{(n)}_0|,|\tilde{Q}^{(n)}_0|}^{(n)}
\end{array}\right],
\end{split}
\end{equation}
and we have
\begin{equation}
\begin{split}
P_{Q}(\boldsymbol{x},q)
&=\left.P_{\tilde{Q}^{(n)}}(\tilde{\boldsymbol{x}}^{(n)},q)\right|_{\tilde x^{(n)}_i=\tilde x^{(n)}_i(x_1,\dots,x_{|Q_0|},q)}\\
&=\left.P_{Q^{(n)}}(\boldsymbol{x}^{(n)},q)\right|_{ x^{(n)}_i= x^{(n)}_i(x_1,\dots,x_{|Q_0|},q)}+O(\boldsymbol{x}^{n+1}).    
\end{split}
\end{equation}
Moreover, from the~previous considerations we know that the~structure
of lowest degrees of contributions to $P_{Q^{(n)}}$ for the~entries
of $\tilde{C}^{(n)}$ reads
\[
\lowest(\tilde{C}^{(n)})=\left[\begin{array}{cccccccc}
(1) &  &  & (n+2) & \cdots & (n+2) & (n+3) & \cdots\\
 & \ddots &  & \vdots & \ddots & \vdots & \vdots\\
 &  & (n) & (2n) & \cdots & (2n) & (2n+2) & \cdots\\
(n+2) & \cdots & (2n) & (n+1) & \cdots & (2n+2) & (2n+3) & \cdots\\
\vdots & \ddots & \vdots & \vdots & \ddots & \vdots & \vdots\\
(n+2) & \cdots & (2n) & (2n+2) & \cdots & (n+1) & (2n+3) & \cdots\\
(n+3) & \cdots & (2n+2) & (2n+3) & \cdots & (2n+3) & (n+2) & \cdots\\
\vdots &  & \vdots & \vdots &  & \vdots & \vdots & \ddots
\end{array}\right].
\]
Therefore, in the~next step we unlink all positive entries and link all negative entries of~$\tilde{C}^{(n)}$ which are above diagonal terms of lowest degree $n+1$. We denote the
resulting quiver~$\tilde{Q}^{(n+1)}$ and the~change of variables $\tilde x^{(n+1)}_i=\tilde x^{(n+1)}_i(x_1,\dots,x_{|Q_0|},q)$. Since unlinking and linking preserve the~quiver motivic generating series, we know that
\begin{equation}
P_{Q}(\boldsymbol{x},q)=\left.P_{\tilde{Q}^{(n+1)}}(\tilde{\boldsymbol{x}}^{(n+1)},q)\right|_{\tilde x^{(n+1)}_i=\tilde x^{(n+1)}_i(x_1,\dots,x_{|Q_0|},q)}.
\end{equation}
If we have $l$ terms of lowest degree $n+1$, the~adjacency matrix
of $\tilde{Q}^{(n+1)}$ is given by
\begin{equation}
\tilde{C}^{(n+1)}=\left[\begin{array}{cccccccc}
\tilde{C}_{11}^{(n)} & 0 & \cdots & 0 & \tilde{C}_{1,k+l+1}^{(n)} & \cdots & \tilde{C}_{1,|\tilde{Q}^{(n)}_0|}^{(n)} & \cdots\\
0 & \tilde{C}_{22}^{(n)} & \ddots & \vdots & \tilde{C}_{2,k+l+1}^{(n)} & \cdots & \tilde{C}_{2,|\tilde{Q}^{(n)}_0|}^{(n)} & \cdots\\
\vdots & \ddots & \ddots & 0 & \vdots & \ddots & \vdots\\
0 & \cdots & 0 & \tilde{C}_{k+l,k+l}^{(n)} & \tilde{C}_{k+l,k+l+1}^{(n)} & \cdots & \tilde{C}_{k+l,|\tilde{Q}^{(n)}_0|}^{(n)} & \cdots\\
\tilde{C}_{1,k+l+1}^{(n)} & \tilde{C}_{2,k+l+1}^{(n)} & \cdots & \tilde{C}_{k+l,k+l+1}^{(n)} & \tilde{C}_{k+l+1,k+l+1}^{(n)} & \cdots & \tilde{C}_{k+1,|\tilde{Q}^{(n)}_0|}^{(n)} & \cdots\\
\vdots & \vdots & \ddots & \vdots & \vdots & \ddots & \vdots\\
\tilde{C}_{1,|\tilde{Q}^{(n)}_0|}^{(n)} & \tilde{C}_{2,|\tilde{Q}^{(n)}_0|}^{(n)} & \cdots & \tilde{C}_{k+l,|\tilde{Q}^{(n)}_0|}^{(n)} & \tilde{C}_{k+1,|\tilde{Q}^{(n)}_0|}^{(n)} & \cdots & \tilde{C}_{|\tilde{Q}^{(n)}_0|,|\tilde{Q}^{(n)}_0|}^{(n)} & \cdots\\
\vdots & \vdots &  & \vdots & \vdots &  & \vdots & \ddots
\end{array}\right].
\end{equation}
The entries denoted by rightmost and downmost dots come from the~unlinking and linking of all entries of $\tilde{C}^{(n)}$ which are above diagonal terms
of lowest degree $n+1$ (so $\tilde{C}^{(n+1)}$ is finite). Their lowest degrees are $n+2$, $n+3$,
$\ldots$ $2n+2$, so they do not alter $P_{Q^{(n)}}$ up to total
degree $n+1$. The~situation is the~same for the~entries above $\tilde{C}_{k+l+1,k+l+1}^{(n)},\ldots,\tilde{C}_{|\tilde{Q}^{(n)}_0||\tilde{Q}^{(n)}_0|}^{(n)}$
-- their lowest degrees are bigger than $n+2$. Summing, up we can
write
\begin{equation}
\lowest(\tilde{C}^{(n+1)})=\left[\begin{array}{ccccc}
(1) &  &  & (n+3) & \cdots\\
 & \ddots &  & \vdots\\
 &  & (n+1) & (2n+3) & \cdots\\
(n+3) & \cdots & (2n+3) & (n+2) & \cdots\\
\vdots &  & \vdots & \vdots & \ddots
\end{array}\right]
\end{equation}
and from the~analysis above we know that the~lowest degree is non-decreasing
when we move right or down in $\tilde{C}^{(n+1)}$.
Let us denote the~$(k+l)\times(k+l)$ top-left corner of
matrix~$\tilde{C}^{(n+1)}$ as 
\begin{equation}
C^{(n+1)}=\left[\begin{array}{cccc}
\tilde{C}_{11}^{(n)} & 0 & \cdots & 0\\
0 & \tilde{C}_{22}^{(n)} & \ddots & \vdots\\
\vdots & \ddots & \ddots & 0\\
0 & \cdots & 0 & \tilde{C}_{k+l,k+l}^{(n)}
\end{array}\right],
\end{equation}
and the~change of variables as $x^{(n+1)}_i=x^{(n+1)}_i(x_1,\dots,x_{|Q_0|},q)$.
Then, we know that for the~corresponding finite diagonal quiver $Q^{(n+1)}$ we have 
\begin{equation}
P_{Q}(\boldsymbol{x},q)=\left.P_{Q^{(n+1)}}(\boldsymbol{x}^{(n+1)},q)\right|_{ x^{(n+1)}_i= x^{(n+1)}_i(x_1,\dots,x_{|Q_0|},q)}+O(\boldsymbol{x}^{n+2})~,
\end{equation}
so $(Q^{(n+1)},\{x^{(n+1)}_i=x^{(n+1)}_i(x_1,\dots,x_{|Q_0|},q)\}_{i\in Q^{(n)}_0})$ is an~$n+1$-st degree approximation of the~initial quiver $Q$.

Since we checked the~theorem for $n=1$ and proved the~induction step,
we know that it is valid for any $n\in\mathbb{Z_{+}}$.
\end{proof}

Let us formalize the~criteria that we used in the~proof above, since they will be important in other constructions:

\begin{dfn}
\label{def:Rules of diagonalization}
We say that the~sequence of linkings and unlinkings follows the~rules of diagonalization if:
\begin{itemize}
    \item For each positive non-diagonal entry $C_{ij}$ we apply unlinking until nodes $i$ and $j$ are disconnected and $C_{ij}=0$.
    \item For each negative non-diagonal entry $C_{ij}$ we apply linking until nodes $i$ and $j$ are disconnected and $C_{ij}=0$.
    \item We unlink and link nodes of lowest degree $n+1$ only if all nodes of lowest degree $n$ are disconnected, i.e. $C^{(n)}$ has already been constructed.
\end{itemize}
\end{dfn}

\subsection{Infinite diagonal quiver --  proof of Theorem \ref{thrm:main}}

Having discussed the~finite approximations of $Q$, we are ready for a~jump to infinity.

\begin{dfn}
\label{def:Qinfinity}
    For any symmetric extended quiver $Q$, the~recursive construction described in the~proof of Theorem \ref{thm:Qn-approx} enables us to construct an~infinite set of pairs
    \begin{equation}
    \{(Q^{(n)},\{x^{(n)}_i=x^{(n)}_i(x_1,\dots,x_{|Q_0|},q)\}_{i\in Q^{(n)}_0})\}_{n\in \mathbb{Z_{+}}},
    \end{equation}
    such that $(Q^{(n)},\{x^{(n)}_i=x^{(n)}_i(x_1,\dots,x_{|Q_0|},q)\}_{i\in Q^{(n)}_0})$ is the~$n$-th degree approximation of $Q$ and
    \begin{align}
        Q^{(n)} & \subseteq Q^{(m)}, \\
        \{x^{(n)}_i=x^{(n)}_i(x_1,\dots,x_{|Q_0|},q)\}_{i\in Q^{(n)}_0} & \subseteq \{x^{(m)}_i=x^{(m)}_i(x_1,\dots,x_{|Q_0|},q)\}_{i\in Q^{(m)}_0}
    \end{align}
    for all $m\geq n$.\footnote{We treat the~superscripts $(n)$ and $(m)$ as decorations which make the expressions more readable, but should not interfere with the~subset relation. For example, we treat sets $\{x^{(2)}_n = q^{-1}x_a x_b \}$ and $\{x^{(3)}_n = q^{-1}x_a x_b \}$ as equal, so $\{x^{(2)}_n = q^{-1}x_a x_b \}\subseteq \{x^{(3)}_n = q^{-1}x_a x_b \}$ is true.}
    
The extended quiver $Q^{(\infty)}$ is a~union of all extended quivers approximating $Q$:
    \begin{equation}\label{eq:infinite union}
        Q^{(\infty)} = \bigcup_{n=1}^{\infty} Q^{(n)},
    \end{equation}
    and the~corresponding identification of the~generating parameters is a~union of all sets of equations:
    \begin{equation}
    \label{eq:infinite identification}
        \{x^{(\infty)}_i=x^{(\infty)}_i(x_1,\dots,x_{|Q_0|},q)\}_{i\in Q^{(\infty)}_0} = \bigcup_{n=1}^{\infty} \{x^{(n)}_i=x^{(n)}_i(x_1,\dots,x_{|Q_0|},q)\}_{i\in Q^{(n)}_0}.
    \end{equation}
\end{dfn}
We will call $Q^{(\infty)}$ the~infinite diagonal quiver. From the~definition it is clear that the~infinite quantity is strictly speaking the~degree of approximation, however in generic case the~size of $Q^{(\infty)}$ is also infinite. Exceptions are described in Sec. \ref{sec:simple}.

Now we are ready to prove the~main theorem of this paper, stating that the~motivic generating series of $Q$ and $Q^{(\infty)}$ are equal upon the~identification of generating parameters given by \eqref{eq:infinite identification}.

\begin{proof}[Proof of Theorem \ref{thrm:main}]
    Assume that        $P_{Q}(\boldsymbol{x},q) - \left.P_{Q^{(\infty)}}(\boldsymbol{x^{(\infty)}},q)\right|_{x^{(\infty)}_i = x^{(\infty)}_i(x_1,\dots,x_{|Q_0|},q)}$ is not zero. Then it is proportional to $\boldsymbol{x}^{\boldsymbol{d}}=\prod_{i=1}^{|Q_0|}x_i^{d_i}$ for some $d_i\in \mathbb{N}$.
    However, we know that $(Q^{(\infty)},\{x^{(\infty)}_i=x^{(\infty)}_i(x_1,\dots,x_{|Q_0|},q)\}_{i\in Q^{(\infty)}_0})$ is a~$|\boldsymbol{d}|$-th degree approximation of $Q$, since
    \begin{align}
        Q^{(|\boldsymbol{d}|)} & \subseteq Q^{(\infty)}, \\
        \{x^{(|\boldsymbol{d}|)}_i=x^{(|\boldsymbol{d}|)}_i(x_1,\dots,x_{|Q_0|},q)\}_{i\in Q^{(|\boldsymbol{d}|)}_0} & \subseteq \{x^{(\infty)}_i=x^{(\infty)}_i(x_1,\dots,x_{|Q_0|},q)\}_{i\in Q^{(\infty)}_0},
    \end{align}
    and for all $i\in Q^{(\infty)}\backslash Q^{(|\boldsymbol{d}|)}$ the~total degree of $x^{(\infty)}_i$ is at least $|\boldsymbol{d}|+1$. Therefore, 
    \begin{equation}
        P_{Q}(\boldsymbol{x},q) - \left.P_{Q^{(\infty)}}(\boldsymbol{x^{(\infty)}},q)\right|_{x^{(\infty)}_i = x^{(\infty)}_i(x_1,\dots,x_{|Q_0|},q)} = O(\boldsymbol{x}^{|\boldsymbol{d}|+1})~,
    \end{equation}
    which is a~contradiction.
\end{proof}

\subsection{Uniqueness}

In this section we show that if the~adjacency matrix of the~initial quiver $Q$ satisfies either $C_{ij}\geq 0$ or $C_{ij}\leq 0$, for all $i,j=1\dots |Q_0|$, then the~entries in the~diagonal quiver matrices $C^{(n)}$ are unique, as long as they satisfy the~rules of diagonalization from Def. \ref{def:Rules of diagonalization}.

\begin{lma}
For any proper or negative quiver $Q$ and $n\in \mathbb{Z}_+$ the~$n$-th approximate quiver $Q^{(n)}$ obtained following the~rules of diagonalization (Def. \ref{def:Rules of diagonalization}) is unique (up to permutation corresponding to relabeling of nodes).
\end{lma}

\begin{proof}
We use the~induction in $n$ and denote the~nodes of $Q^{(n)}$ for any $n$ as $i\in \{1,\dots,k\}$, keeping in mind that $Q_0^{(n)} \subseteq Q_0^{(n+1)}$. Without loss of generality we focus on the~case of $C_{ij}\geq 0$.
Given a~symmetric quiver $Q$,
its $n$-th approximate quiver $Q^{(n)}$ by definition satisfies
\begin{equation}
    P_Q(\boldsymbol{x},q) - P_{Q^{(n)}}(\boldsymbol{x},q) = O(\boldsymbol{x}^{n+1})\,.
\end{equation}
Alternatively, this can be written in terms of DT invariants as
\begin{equation}
    \Omega_Q(\boldsymbol{x},q) - \Omega_{Q^{(n)}}(\boldsymbol{x},q) = O(\boldsymbol{x}^{n+1})\,.
\end{equation}
We have
\begin{equation}
C^{(1)} =
\renewcommand{\arraystretch}{1.75}
\left[
\begin{array}{ccc}
    C_{11}  & & \\
    &  \ddots & \\
    &  & C_{|Q_0||Q_0|} 
\end{array}
\right]~,
\qquad
C^{(n)} =
\left[
\begin{array}{ccc}
    C_{11}  & & \\
    &  \ddots & \\
    & & C_{kk}\\
\end{array}
\right]~,
\end{equation}
where $k>|Q_0|$ and blank space denotes zero entries.
The first degree approximation to $Q$ is defined from the~main diagonal of $C$, therefore $C^{(1)}$
is manifestly unique.
We further assume that $Q^{(n)}$
is also unique. For $(n+1)$ we write
\begin{equation}
C^{(n+1)} =
\left[
\begin{array}{ccc|ccc}
    C_{11}  & & & & & \\
    &  \ddots & & & & \\
    &  & C_{kk} & & & \\
    \hline
    &  & & C_{k+1,k+1} & & \\
    &  & & & \ddots & \\
    &  & & & & C_{k+l,k+l}
\end{array}
\right] = 
\left[
\begin{array}{ccc|ccc}
    &  & & & & \\
    &  C^{(n)} & & & & \\
    &  &  & & & \\
    \hline
    &  & &  & & \\
    &  & & & R^{(n)} & \\
    &  & & & & \\
\end{array}
\right]\, ,
\end{equation}
%
where the~first block is just $C^{(n)}$ and the~rest is a~quiver of size $l$, which we denoted $R^{(n)}$, so that
\begin{equation}
\Omega_{Q^{(n+1)}}(\boldsymbol{x},q)= \Omega_{Q^{(n)}}(\boldsymbol{x},q) +  \Omega_{R^{(n)}}(\boldsymbol{x},q)\,.
\end{equation}
$Q^{(n)}$ is uniquely fixed by induction assumption and it contains all nodes with the~lowest degree $\leq n$. The~contributions from $R^{(n)}$ to $P_Q$ are of degree $\geq n+1$, but since we are interested only in the~$n+1$-st approximation, only the~lowest degree contributions matter.
Therefore, we need to check whether it is possible to have $R'(n)\neq R(n)$ such that lowest degree terms of $\Omega_{R'(n)}(\boldsymbol{x},q)$ are equal to those of $\Omega_{R(n)}(\boldsymbol{x},q)$.
To clarify that, we study the~coefficient in front of the~monomial $\boldsymbol{x}^I=x_1^{i_1}\dots x_{|Q_0|}^{i_{|Q_0|}}$ of total degree $n+1$ (the lowest degree) in $\Omega_{R^{(n)}}(\boldsymbol{x},q)$.
Let us assume that this monomial arises from the~change of variables associated to the~quiver parameters $x_{j_1},\dots, x_{j_s}$ in $R^{(n)}$. Since ${x}^I$ is of total degree $n+1$, each $x_{j_r}$ ($r\in \{1,\dots,s\}$) originated from $n$ unlinkings and (\ref{eq:unlinking change of vars}) implies that $x_{j_r}=q^{-n}\boldsymbol{x}^I$.
On the~other hand, the~generating series of DT invariants of each $C_{j_rj_r}$-loop quiver reads (\ref{eq:m-loop DT})
\begin{equation}\label{eq:DT Rn loop}
    \Omega_{\text{$C_{j_rj_r}$-loop}}(x_{j_r},q) = (-1)^{C_{j_rj_r}-1}q^{C_{j_rj_r}}x_{j_r} + O(\boldsymbol{x}^{n+2})\,.
\end{equation}
Substituting $x_{j_r}=q^{-n}\boldsymbol{x}^I$ and summing over $r$ we obtain the~explicit form of the~coefficient:
%
\begin{equation}\label{eq:Omega coeff}
    \coeff_{\boldsymbol{x}^I} \Omega_{R^{(n)}}(\boldsymbol{x},q) = q^{-n}\,((-1)^{C_{j_1j_1}-1}q^{C_{j_1j_1}}+\ldots +(-1)^{C_{j_sj_s}-1}q^{C_{j_sj_s}})\,.
\end{equation}
It is now evident that the~only allowed operations which preserve the~coefficient (\ref{eq:Omega coeff}) are permutations of the~diagonal entries and thus the~quiver $R^{(n)}$ is unique. This completes the~induction step, and the~reasoning for linking negative quivers follows exactly the~same path. Note that if linking and unlinking are combined, this argument does not work, since then we do not get the~overall factor (like $q^{-n}$ in \eqref{eq:Omega coeff}) and one could smartly trade the~number of linkings, unlinkings and loops to get the~same power of $q$.
\end{proof}

We can easily generalize this result to approximations of an~infinite degree.

\begin{thm}\label{thm:Uniqueness of Q inf}
For any proper or negative quiver $Q$ the~infinite diagonal quiver $Q^{(\infty)}$ obtained following the~rules of diagonalization (Def. \ref{def:Rules of diagonalization}) is unique (up to permutation corresponding to relabeling of nodes).
\end{thm}

\begin{proof}
Given that $Q^{(n)}$ is unique for every $n\in \mathbb{Z}_+$, we make use of the~definition of $Q^{(\infty)}$ via set-theoretic union (\ref{eq:infinite union}), taking into account identifications (\ref{eq:infinite identification}),
which proves the~statement.
\end{proof}

It remains unclear whether the~infinite diagonal quiver is unique for an~extended quiver with both positive and negative entries. This important question will be taken into account in the~future.
However, since such extended quivers occur frequently in many physical applications, we take a~look at a~few of them in Sec. \ref{sec:KQ correspondence} and \ref{sec:FK quivers}.

\section{DT invariants from infinite diagonal quivers}
\label{sec:DT}

In this section we show how we can apply the~knowledge of infinite diagonal quivers to obtain a~novel way of analyzing DT invariants. This framework leads to a~proof of integrality of DT invariants for any extended quiver, as well as interesting physical and algebraic interpretations.

\subsection{New proof of integrality of DT invariants}
\label{sec:Integrality}

At this stage of our journey we have discovered two important facts:
\begin{itemize}
    \item The~motivic generating series of any $m$-loop quiver can be factorized into $q$-Pochhammers with integer powers (Lemma \ref{lma:mloop} combined with (\ref{eq:m-loop-factorized})).
    \item The~motivic generating series of any symmetric extended quiver can be factorized into an~infinite product of the~motivic generating series of $m$-loop quivers (Theorem~\ref{thrm:main}). 
\end{itemize}
Combining the~above statements we arrive at

\begin{thm}
\label{thm:KS}
For any extended symmetric quiver $Q$ (see Def. \ref{def:extended quiver}), its motivic DT invariant admits the~following decomposition:
\begin{equation}\label{eq:Omegas-from-m-loop-quivers}
    \Omega_Q(x_1,\dots,x_{|Q_0|},q) = \sum_{i\in Q^{(\infty)}_0}
    \Omega_{\text{$C^{(\infty)}_{ii}$-loop}}(x^{(\infty)}_i(x_1,\dots,x_{|Q_0|}),q)\, .
\end{equation}
Moreover, if $Q$ is a~proper or negative quiver, then this decomposition is unique.
\end{thm}

\begin{proof}
Theorem \ref{thrm:main} tells us that for every formal symmetric quiver $Q=(Q_0,Q_1)$ there exists an~infinite diagonal quiver $Q^{(\infty)}$ such that $P_Q=P_{Q^{(\infty)}}$. Therefore, motivic generating series for $Q$ can be written as a~product of $m$-loop quiver series:
\begin{equation}\label{eq:infinite_factorization}
    P_Q(\boldsymbol{x},q) = \prod_{i\in Q^{(\infty)}_0}
    \left.P_{\text{$C^{(\infty)}_{ii}$-loop}}(x^{(\infty)}_i,q)\right|_{x^{(\infty)}_i=x^{(\infty)}_i(x_1,\dots,x_{|Q_0|},q)}~.
\end{equation}
Each term in the~above product corresponds to a~diagonal entry in $C^{(\infty)}$, and we can express motivic DT invariants of $Q$ as a~linear combination of $m$-loop quiver invariants, which yields the~statement (as uniqueness follows from Theorem  \ref{thm:Uniqueness of Q inf}).
\end{proof}

\begin{cor}
For any extended quiver $Q$, its motivic DT invariants are integer.
\end{cor}

\begin{proof}
It is left to apply the~results from Sec. \ref{subsec:m-loop}, which, combined with Eq. (\ref{eq:Omegas-from-m-loop-quivers}), confirm integrality of $\Omega_Q$
in terms of DT invariants for $m$-loop quivers. 
\end{proof}

An important implication of Theorem \ref{thm:KS} is that the~DT invariants for an~arbitrary symmetric quiver obtain a~purely combinatorial description.
Indeed, Eq. (\ref{eq:Omegas-from-m-loop-quivers}) shows that motivic DT invariants of $Q$ are essentially captured by the~construction of the~DT invariants for $m$-loop quiver given in \cite{Rei12,KS1608} along with the~data coming from unlinking and linking. 
In practice, $\Omega_{\boldsymbol{d},s}$ 
for a~given vector $\boldsymbol{d}$ can be computed in two steps:

\begin{itemize}
    \item Finding the~$|\boldsymbol{d}|$-th approximation quiver $Q^{(|\boldsymbol{d}|)}$ (see Sec. \ref{sec:nth degree approximation}), which is diagonal by construction, and whose motivic generating series are given by a~product of $m$-loop quivers.
    \item Computing the~motivic DT invariants $\Omega_Q^{(|\boldsymbol{d}|)}(x_1,\dots,x_{|Q_0|},q)$ for $Q^{(|\boldsymbol{d}|)}$ using the~formulas for $m$-loop quivers, .
\end{itemize}

Note that the~definition of approximate quiver implies that

\begin{equation}\label{eq:Omegas-from-approximations}
    \Omega_Q(x_1,\dots,x_{|Q_0|},q) - 
    \Omega_Q^{(|\boldsymbol{d}|)}(x_1,\dots,x_{|Q_0|},q) = O(\boldsymbol{x}^{\boldsymbol{d}+1})\,.
\end{equation}

This allows to identify $\Omega_{\boldsymbol{d}',s}=\Omega^{(|\boldsymbol{d}|)}_{\boldsymbol{d}',s}$
for any monomial of total degree $|\boldsymbol{d}'|\leq |\boldsymbol{d}|$.
In particular, since $Q^{(|\boldsymbol{d}|)}$ is a~finite diagonal quiver, there are only finitely many DT invariants $\Omega_{\boldsymbol{d}',s}\neq 0$ for a~fixed $\boldsymbol{d}'$.
One can then use the~recurrence relations of Ref.~\cite{KS1608} to directly compute the~$C_{ii}^{(|\boldsymbol{d}|)}-$loop quivers invariants. Afterwards, one unfolds changes of variables used in (un)linkings, and identifies the~desired DT invariants in order to find $\Omega_Q^{(|\boldsymbol{d}|)}$ and thus $\Omega_Q$ up to a~given order in $x_1,\dots,x_{|Q_0|}$.
We discuss these steps in detail in examples in Sec.~\ref{sec:simple}.

Summing up, the~above procedure allows for an~effective computational approach for the~DT invariants of any symmetric quiver. As a~useful application one can resort to the~knots-quivers correspondence \cite{KRSS1707short,KRSS1707long} and compute the~LMOV invariants of knots, see Sec. \ref{sec:KQ correspondence}. Another application, which 
uses the~fact that quivers can be related to the~$F_K$~invariants of knot complements \cite{Kuch2005}, is discussed in Sec. \ref{sec:FK quivers}.

\subsection{Physical and algebraic interpretations}
\label{sec:gauge}

Let us study the~physical and algebraic interpretations of the~results obtained in previous sections. 

Following \cite{EKL1811}, we know that to every symmetric quiver $Q$ we can assign a~$3$d $\mathcal{N}=2$ gauge theory $T[Q]$ using the~following rules:
\begin{itemize}
    \item Gauge group: $U(1)^{(1)}\times\dots\times U(1)^{(|Q_0|)}$
    \item Matter: chiral fields $\Phi_i$ charged $\delta_{ij}$ under $U(1)^{(j)}$
    \item Chern-Simons (CS) couplings: $\kappa^{\rm eff}_{ij}=C_{ij}$
    \item Fayet-Ilioupoulos (FI) couplings: $\zeta_{i}=\log\left(\left(-1\right)^{C_{ii}}x_i\right)$
\end{itemize}
To that end, an~$m$-loop quiver corresponds to a~$U(1)$ level $m$ Chern-Simons gauge theory coupled to a~single chiral superfield. More concretely, the~partition function of this theory is equal to the~motivic generating series $P_{m\text{-loop}}(x,q)$ after appropriate identification of variables, as shown in \cite{EKL1811}.

Theorem \ref{thrm:main} implies that with appropriate choice of levels and FI couplings (corresponding to infinite diagonal quiver data), the~product of simple $U(1)$ Chern-Simons theories described above can reproduce the~partition function of theory $T[Q]$ for any symmetric quiver $Q$. In other words, the~theory $T[Q^{(\infty)}]$ corresponding to infinite diagonal quiver is dual to $T[Q]$. It is possible because the~whole structure encoded in the~CS couplings connecting different $U(1)$ theories (corresponding to quiver arrows) is translated into FI couplings (corresponding to changes of variables). 

The duality between $T[Q]$ and $T[Q^{(\infty)}]$ implies the~equality of their BPS spectra, however they are built in different ways. In case of $Q$, we have basic states corresponding to quiver nodes and the~their bound states coming from interactions encoded in the~quiver arrows \cite{EKL1811}. On the~other hand, the~BPS spectrum of $Q^{(\infty)}$ consists only of basic states, which are non-interacting from the~perspective of CS couplings, but become interdependent via FI couplings.
One can also view $T[Q]$ and $T[Q^{(\infty)}]$ as two ends of the~chain of dualities between $3$d $\mathcal{N}=2$ theories. Each duality corresponds to unlinking or linking one pair of nodes \cite{EKL1910}. Physically, it means that one bound state and the~corresponding CS coupling are translated into a~basic state and the~FI coupling. Since the~BPS spectrum of $T[Q]$ is usually infinite, so is the~chain of dualities coming from unlinking and liniking. However, we can group these dualities into two-level structure. 

The first level comes from the~fact that whenever we have $C_{ij}>1$ pairs of arrows connecting nodes $i$ and $j$ with $C_{ii}$ and $C_{jj}$ loops respectively, then unliking leads to a~new node with $C_{ii}+C_{jj}+2C_{ij}-1$ loops and a~corresponding quiver variable given by $q^{-1}x_i x_j$. However, we still have $C_{ij}-1>0$ pairs of arrows to unlink. The~change of variables remains the~same, but now there are $C_{ii}+C_{jj}+2C_{ij}-3$ loops at the~new node. In consequence, unlinking $C_{ij}>1$ leads to a~sequence of $C_{ij}$ new nodes with the~number of loops following the~double factorial pattern:
\begin{equation}
C_{ii}+C_{jj}+2C_{ij}-1,\quad C_{ii}+C_{jj}+2C_{ij}-3,\quad C_{ii}+C_{jj}+2C_{ij}-5,\,\dots,\,  C_{ii}+C_{jj}+1
\end{equation}
(the corresponding quiver variables are all given by $q^{-1}x_i x_j$). In analogy, linking $C_{ij}<-1$ leads to a~sequence of $|C_{ij}|$ new nodes with the~number of loops following the~double factorial pattern:
\begin{equation}
C_{ii}+C_{jj}+2C_{ij},\quad C_{ii}+C_{jj}+2C_{ij}+2,\quad C_{ii}+C_{jj}+2C_{ij}+4,\,\dots,\,  C_{ii}+C_{jj}-2
\end{equation}
(the corresponding quiver variables are all given by $x_i x_j$).

The second level of the~structure of the~chain of dualities is given by the~lowest degree of quiver nodes. We start from $Q$ where all nodes are of lowest degree $1$. After unlinking (or linking) them, we obtain nodes of lowest degree $2$. Unlinking the~arrows connecting the~new nodes with the~old ones produces nodes of lowest degree $3$, and so on.
This grouping is clearly visible in the~proof of Theorem \ref{thm:Qn-approx}, since the~difference between $Q^{(n)}$ and $Q^{(n+1)}$ lies exactly in nodes of lowest degree $n+1$. 
We will see these two levels of the~structure of the~chain of dualities in various examples of quivers in the~following sections.

Finally, we mention the~intriguing possibility for an~algebraic interpretation of Theorem~\ref{thm:KS}.
In \cite{DFR2111}, certain Lie superalgebra $\mathfrak{g}_Q$, whose Koszul dual is presumably related to Cohomological Hall algebra of Kontsevich and Soibelman \cite{KS0811}, has been defined for every symmetric quiver.
$\mathfrak{g}_Q$ has generators
$b_{i,k}$ for $i=1\dots |Q_0|$, subject to relations for all $i,j=1\dots |Q_0|$ and $k,l\in\mathbb{N}$:
\begin{align}
    [b_{i,k},b_{i,l}] &= 0, \quad C_{ii} = 0, \nonumber\\
    \sum_{p=0}^{C_{ii}-1}(-1)^p\, \binom{C_{ii}-1}{p}\,[b_{i,k-p},b_{i,l+p}] &= 0,\quad C_{ii}\geq 1, \\
    \sum_{p=0}^{C_{ij}}(-1)^p\, \binom{C_{ij}}{p}\,[b_{i,k-p},b_{j,l+p}] &= 0,\quad i\neq j. \nonumber
\end{align}

The main outcome of this construction is that the~Poincare series for $\mathfrak{g}_Q$ coincides with the~motivic DT invariant of $Q$, up to an~overall factor. In this context (\ref{eq:Omegas-from-m-loop-quivers}) suggests that $\mathfrak{g}_Q$ admits a~decomposition into ``elementary'' Lie algebras associated to $m$-loop quivers. Furthermore, we expect a~relation between (un)linking and the~Lie bracket. These considerations deserve a~thorough study, especially in relation to negative quivers (for example, in the~context of the~vertex operator description in \cite{DFR2111}), and we leave it for future work.

\section{Simple examples}
\label{sec:simple}

In the~following part of the~paper we illustrate our results in explicit examples. To start with, in this section we consider simple quivers with one or two pairs of arrows connecting different nodes. In the~next sections we discuss examples related to the~knots-quivers correspondence. 

While examples considered in this paper are relatively uncomplicated and we are able to analyze them explicitly, in analysis of more intricate cases one could take advantage of prequivers~\cite{JKLNS2105}, which encode a~part of the~spectrum and simplify computations by dividing it in steps.\footnote{The idea of this approach follows from the~observation that (un)linking and splitting of prequiver nodes commute. One way to understand this is to notice that unlinking does not affect the~diagonal terms, while splitting acts solely on them. }

\subsection{Symmetric \texorpdfstring{$A_2$}{A2}
quiver}

The simplest example of the~quiver diagonalization corresponds to the~unlinking of the~single pair of arrows in the~symmetrization of the~$\vec{A}_2$ Dynkin quiver. The~quiver and its adjacency matrix are given by

\begin{equation}
\label{eq:A2quiver}
Q= \begin{tikzcd}
	{\bullet} & {\bullet}
	\arrow[curve={height=-6pt}, from=1-1, to=1-2]
	\arrow[curve={height=6pt}, tail reversed, no head, from=1-1, to=1-2]~,
\end{tikzcd}
\qquad
C=
\begin{bmatrix}
0 & 1 \\
1 & 0
\end{bmatrix},
\end{equation}
and we will call it a~symmetric $A_2$ quiver.

Let us follow Sec. \ref{sec:nth degree approximation} and construct all $n$-th degree approximations of $Q$. From \eqref{eq:1st order approx} we immediately find that
\begin{equation}
\label{eq:A2quiver 1st order}
C^{(1)}=
\begin{bmatrix}
0 & 0 \\
0 & 0
\end{bmatrix}
=\diag(0,0),
\qquad
x^{(1)}_1 = x_1,
\quad
x^{(1)}_2 = x_2,
\end{equation}
and Eq.~\eqref{eq:PQ for 1st order approx} takes a~very simple form:
\begin{equation}
    P_{Q}(\boldsymbol{x},q)=1+\frac{x_{1}}{1-q^{2}}+\frac{x_{2}}{1-q^{2}}+O(\boldsymbol{x}^{2}).
\end{equation}
We can obtain the~matrix $\tilde C^{(1)}$ and the~change of variables $\tilde x^{(1)}_i=\tilde x^{(1)}_i(x_1,x_2,q)$ by applying the~unlinking operation described in Def. \ref{def:(un)linking} and Theorem \ref{thm:(un)linking}:
\begin{equation}
\label{eq:A2quiver unlinked}
\tilde C^{(1)}=
\begin{bmatrix}
0 & 0 & 0\\
0 & 0 & 0 \\
0 & 0 & 1
\end{bmatrix}=\diag(0,0,1),
\qquad
\tilde x^{(1)}_1 = x_1,
\quad
\tilde x^{(1)}_2 = x_2,
\quad
\tilde x^{(1)}_3 = q^{-1}x_1 x_2.
\end{equation}
We can see that $Q^{(1)}$ is a~diagonal subquiver of $\tilde Q^{(1)}$ and the~lowest degrees of contributions to $P_{\tilde{Q}^{(1)}}$ for entries of $\tilde C^{(1)}$ are given by
\begin{equation}
\text{lowest\_degree}(\tilde{C}^{(1)})=
\begin{bmatrix}
(1) & (2) & (3) \\
(2) & (1) & (3) \\
(3) & (3) & (2)
\end{bmatrix}.
\end{equation}
In the~next step, we would normally unlink the~new node from the~initial ones, but we can see that $\tilde Q^{(1)}$ is already diagonal. In consequence, we have
\begin{equation}
\label{eq:A2quiver all n}
C^{(n)}=
\begin{bmatrix}
0 & 0 & 0\\
0 & 0 & 0 \\
0 & 0 & 1
\end{bmatrix}=\diag\left(0,0\dashline 1\right),
\qquad
x^{(n)}_1 = x_1,
\quad
x^{(n)}_2 = x_2,
\quad
x^{(n)}_3 = q^{-1}x_1 x_2
\end{equation}
for all $n>1$, where the~dashed line separates the~nodes which belong to $n$-th and $n+1$-st approximations.
Following the~Def. \ref{def:Qinfinity}, it leads to 
\begin{equation}
\label{eq:A2 infinite quiver}
C^{(\infty)}=
\begin{bmatrix}
0 & 0 & 0\\
0 & 0 & 0 \\
0 & 0 & 1
\end{bmatrix}=\diag\left(0,0 \dashline 1\right),
\qquad
x^{(\infty)}_1 = x_1,
\quad
x^{(\infty)}_2 = x_2,
\quad
x^{(\infty)}_3 = q^{-1}x_1 x_2.
\end{equation}
Having found $Q^{(\infty)}$, we can construct the~BPS spectrum of $Q$ using our knowledge about $m$-loop quivers, described in Sec.~\ref{sec:DT}. In the~case of symmetric $A_2$ quiver,  Eq.~\eqref{eq:infinite_factorization} takes form
\begin{equation}\label{eq:A2_product_form}
    P_Q(\boldsymbol{x},q) =
    P_{\text{$0$-loop}}(x_1,q)
    P_{\text{$0$-loop}}(x_2,q)
    P_{\text{$1$-loop}}(q^{-1}x_1 x_2,q).
\end{equation}
From Tab.~\ref{tab:mqPoch} we can immediately read off that \begin{equation}
    P_Q(\boldsymbol{x},q) =
    (x_1;q^2)^{-1}_{\infty}
    (x_2;q^2)^{-1}_{\infty}
    (x_1 x_2;q^2)^{+1}_{\infty},
\end{equation}
so DT invariants are given by
\begin{equation}
    \Omega(\boldsymbol{x},q) = -x_1 -x_2 +x_1 x_2 \quad \Leftrightarrow \quad \Omega_{(1,0),0}=-1,\; \Omega_{(0,1),0}=-1,\;\Omega_{(1,1),0}=+1.
\end{equation}

The symmetric $A_2$ quiver is exceptional in a~twofold sense. First, $Q^{(\infty)}$ is finite, which comes from the~fact that the~pair $(Q^{(2)},x^{(2)}_i=x^{(2)}_i(x_1,x_2,q))$ is not only a~second degree approximation of $Q$, but also gives an~exact expression for the~motivic generating series:
\begin{equation}
    P_{Q}(\boldsymbol{x},q)=\left.P_{Q^{(2)}}(\boldsymbol{x}^{(2)},q)\right|_{ x^{(2)}_i= x^{(2)}_i(x_1,x_2,q)}.
\end{equation}
Second, $Q^{(\infty)}$ is not only diagonal, but all nodes have either zero or one loop. This leads to a~finite BPS spectrum consisting of 3 states. In this simple case we can see clearly that after diagonalization the~bound state arising from the~interaction of nodes $1$ and $2$ (represented by the~pair of arrows) becomes a~basic state associated to the~node $3$.

To summarize, the~unlinking data which generate the~infinite diagonal quiver in this example can be compactly presented in the~table form (see Tab. \ref{tab:Cinf_A2}).
\begin{table}[h!]
    \centering
    \renewcommand{\arraystretch}{1.25}
    \begin{tabular}{||c|c||}
        \hline
        $m$ & unlinkings \\
        \hline
        \hline
        0 & $x_1,\ x_2$ \\
        \hline
        1 & $(x_1,x_2)$ \\
        \hline
    \end{tabular}
    \caption{$Q^{(\infty)}$ data for symmetric $A_2$ quiver.}
    \label{tab:Cinf_A2}
\end{table}
The first row encodes two copies of 0-loop quiver, each corresponding to $x_i$; the~second row is one copy of 1-loop quiver and $(x_1,x_2)$ denotes the~node created at unlinking of the~initial pair of nodes. Therefore, by looking at this table one can immediately write down the~factorized form~(\ref{eq:A2_product_form}). This phenomenon is closely related to the~pentagon relation \cite{KS0811}, which was noticed in \cite{EKL1910}. Analogous features can be stated for more complicated quivers, as we will discuss in the~upcoming paper \cite{JKLNS}.

\subsection{Symmetric \texorpdfstring{$A_3$}{A3} quiver}
\label{sec:SimpEx}

The next example we consider is a~symmetrization of Dynkin quiver $\vec{A}_3$: 
\begin{equation}\label{eq:A3quiver}
Q=\begin{tikzcd}
	{\bullet} & {\bullet} & {\bullet}
	\arrow[curve={height=-6pt}, from=1-2, to=1-3]
	\arrow[curve={height=-6pt}, from=1-3, to=1-2]
	\arrow[curve={height=-6pt}, from=1-1, to=1-2]
	\arrow[curve={height=6pt}, tail reversed, no head, from=1-1, to=1-2]~
\end{tikzcd},
\qquad
C=
\begin{bmatrix}
0 & 1 & 0 \\
1 & 0 & 1 \\
0 & 1 & 0 \\
\end{bmatrix},
\qquad
\boldsymbol{x} = \left[
\begin{array}{c}
x_1\\
x_2 \\
x_3 \\
\end{array} \right].
\end{equation}
Unlike the~$\vec{A}_2$ case, here the~set of DT invariants is infinite, and so $Q^{(\infty)}$ has infinitely many nodes. Nevertheless, we can take advantage of the~construction of $n$-th degree approximation quivers $Q^{(n)}$ to compute $Q^{(\infty)}$ up to any given order.

Let us begin with the~first degree approximation. We can consecutively unlink $(x_1,x_2)$ and $(x_2,x_3)$ (pairs of arrows that will be unlinked are highlighted in red) 
\begin{equation}\label{eq:A3_sequencea}
C =
\begin{tikzpicture}[
    node distance=1mm and 0mm,
    baseline]
\matrix (M1) [matrix of nodes,{left delimiter=[},{right delimiter=]}]
{
 0 & \textcolor{red}{1} & 0 \\
 \textcolor{red}{1} & 0 & 1 \\
 0 & 1 & 0 \\
};
\draw[black,dashed] 
        (M1-1-1.north west) -| (M1-3-3.south east) -| (M1-1-1.north west);
\end{tikzpicture}
\ \rightarrow \
\begin{tikzpicture}[
    node distance=1mm and 0mm,
    baseline]
\matrix (M1) [matrix of nodes,{left delimiter=[},{right delimiter=]}]
{
 0 & 0 & 0 & 0 \\
 0 & 0 & \textcolor{red}{1} & 0 \\
 0 & \textcolor{red}{1} & 0 & 1 \\
 0 & 0 & 1 & 1 \\
};
\draw[black,dashed] 
        (M1-1-1.north west) -| (M1-3-3.south east) -| (M1-1-1.north west);
\end{tikzpicture}
\ \rightarrow \
\begin{tikzpicture}[
    node distance=1mm and 0mm,
    baseline]
\matrix (M1) [matrix of nodes,{left delimiter=[},{right delimiter=]}]
{
 0 & 0 & 0 & 0 & 0 \\
 0 & 0 & 0 & 0 & 0 \\
 0 & 0 & 0 & 1 & 0 \\
 0 & 0 & 1 & 1 & 1 \\
 0 & 0 & 0 & 1 & 1 \\
};
\draw[black,dashed] 
        (M1-1-1.north west) -| (M1-3-3.south east) -| (M1-1-1.north west);
\end{tikzpicture}
= \tilde{C}^{(1)}
\,,
\end{equation}
where the~last two nodes coming from unlinking correspond to quiver variables $q^{-1}x_1x_2$ and $q^{-1}x_2x_3$. This gives the~first degree approximation
\begin{equation}
C^{(1)} = 
    \begin{bmatrix}
    0 & 0 & 0 \\
    0 & 0 & 0 \\
    0 & 0 & 0 \\
    \end{bmatrix},
    \qquad x_1^{(1)} = x_1,\quad x_2^{(1)} = x_2,\quad x_3^{(1)} = x_3.
\end{equation}

In order to get the~second degree approximation, we start from $\tilde{C}^{(1)}$ and perform unlinking at $(x_3,(x_1,x_2))$, i.e. we unlink the~node corresponding to $x_3$ and the~node that came from unlinking $x_1$ and $x_2$. In the~second step\footnote{The ordering of the~steps corresponding to the~same degree is arbitrary.} we perform unlinking at $(x_3,(x_1,x_2))$, which gives $\tilde{C}^{(2)}$:
\begin{equation}
\begin{aligned}
\tilde{C}^{(1)} =
\begin{tikzpicture}[
    node distance=1mm and 0mm,
    baseline]
\matrix (M1) [matrix of nodes,{left delimiter=[},{right delimiter=]}]
{
 0 & 0 & 0 & 0 & 0 \\
 0 & 0 & 0 & 0 & 0 \\
 0 & 0 & 0 & \textcolor{red}{1} & 0 \\
 0 & 0 & \textcolor{red}{1} & 1 & 1 \\
 0 & 0 & 0 & 1 & 1 \\
};
\draw[black,dashed] 
        (M1-1-1.north west) -| (M1-5-5.south east) -| (M1-1-1.north west);
\end{tikzpicture}
\ \rightarrow \
\begin{tikzpicture}[
        node distance=1mm and 0mm,
    baseline]
\matrix (M1) [matrix of nodes,{left delimiter=[},{right delimiter=]}]
{
 0 & 0 & 0 & 0 & 0 & 0 \\
 0 & 0 & 0 & 0 & 0 & 0 \\
 0 & 0 & 0 & 0 & 0 & 0 \\
 0 & 0 & 0 & 1 & \textcolor{red}{1} & 1 \\
 0 & 0 & 0 & \textcolor{red}{1} & 1 & 1 \\
 0 & 0 & 0 & 1 & 1 & 2 \\
};
\draw[black,dashed] 
        (M1-1-1.north west) -| (M1-5-5.south east) -| (M1-1-1.north west);
\end{tikzpicture}
\ \rightarrow \
\begin{tikzpicture}[
        node distance=1mm and 0mm,
    baseline]
\matrix (M1) [matrix of nodes,{left delimiter=[},{right delimiter=]}]
{
 0 & 0 & 0 & 0 & 0 & 0 & 0 \\
 0 & 0 & 0 & 0 & 0 & 0 & 0 \\
 0 & 0 & 0 & 0 & 0 & 0 & 0 \\
 0 & 0 & 0 & 1 & 0 & 1 & 1 \\
 0 & 0 & 0 & 0 & 1 & 1 & 1 \\
 0 & 0 & 0 & 1 & 1 & 2 & 2 \\
 0 & 0 & 0 & 1 & 1 & 2 & 3 \\
};
\draw[black,dashed] 
        (M1-1-1.north west) -| (M1-5-5.south east) -| (M1-1-1.north west);
\end{tikzpicture}
= \tilde{C}^{(2)}
\,.
\end{aligned}
\end{equation}
The top-left $5\times 5$ subquiver is diagonal and contains all entries of lowest degree $\leq2$, therefore it provides the~second degree approximation of $Q$:
\begin{equation}
C^{(2)} = 
    \left[
    \begin{array}{ccccc}
    0 & 0 & 0 & 0 & 0 \\
    0 & 0 & 0 & 0 & 0 \\
    0 & 0 & 0 & 0 & 0 \\
    0 & 0 & 0 & 1 & 0 \\
    0 & 0 & 0 & 0 & 1 \\
    \end{array}
    \right],
    \qquad
    \begin{aligned}
    x_1^{(2)} = x_1,\quad x_2^{(2)} = x_2,\quad x_3^{(2)} = x_3,\quad \\ x_4^{(2)} = q^{-1}x_1x_2,\quad x_5^{(2)} = q^{-1}x_2x_3.
    \end{aligned}
\end{equation}

Likewise, to get the~cubic approximation we unlink all nodes of lowest degree $\leq 3$. They are inside the~dashed submatrix of $\tilde{C}^{(2)}$ (the bottom-right entry 3 corresponding to $\tilde x_7^{(2)}$ is of lowest degree 4):
\begin{equation}
\renewcommand{\arraystretch}{1.25}
\begin{tabular}{c}
$\tilde{C}^{(2)} =$
\begin{tikzpicture}[
    node distance=1mm and 0mm,
    baseline]
\matrix (M1) [matrix of nodes,{left delimiter=[},{right delimiter=]}]
{
 0 & 0 & 0 & 0 & 0 & 0 & 0 \\
 0 & 0 & 0 & 0 & 0 & 0 & 0 \\
 0 & 0 & 0 & 0 & 0 & 0 & 0 \\
 0 & 0 & 0 & 1 & 0 & \textcolor{red}{1} & 1 \\
 0 & 0 & 0 & 0 & 1 & 1 & 1 \\
 0 & 0 & 0 & \textcolor{red}{1} & 1 & 2 & 2 \\
 0 & 0 & 0 & 1 & 1 & 2 & 3 \\
};
\draw[black,dashed] 
        (M1-1-1.north west) -| (M1-6-6.south east) -| (M1-1-1.north west);
\end{tikzpicture}
$\ \rightarrow \ $
\begin{tikzpicture}[
    node distance=1mm and 0mm,
    baseline]
\matrix (M1) [matrix of nodes,{left delimiter=[},{right delimiter=]}]
{
 0 & 0 & 0 & 0 & 0 & 0 & 0 & 0 \\
 0 & 0 & 0 & 0 & 0 & 0 & 0 & 0 \\
 0 & 0 & 0 & 0 & 0 & 0 & 0 & 0 \\
 0 & 0 & 0 & 1 & 0 & 0 & 1 & 1 \\
 0 & 0 & 0 & 0 & 1 & \textcolor{red}{1} & 1 & 1 \\
 0 & 0 & 0 & 0 & \textcolor{red}{1} & 2 & 2 & 2 \\
 0 & 0 & 0 & 1 & 1 & 2 & 3 & 3 \\
 0 & 0 & 0 & 1 & 1 & 2 & 3 & 4 \\
};
\draw[black,dashed] 
        (M1-1-1.north west) -| (M1-6-6.south east) -| (M1-1-1.north west);
\end{tikzpicture}
$\ \rightarrow \ $
\\
\begin{tikzpicture}[
    node distance=1mm and 0mm,
    baseline]
\matrix (M1) [matrix of nodes,{left delimiter=[},{right delimiter=]}]
{
 0 & 0 & 0 & 0 & 0 & 0 & 0 & 0 & 0 \\
    0 & 0 & 0 & 0 & 0 & 0 & 0 & 0 & 0 \\
    0 & 0 & 0 & 0 & 0 & 0 & 0 & 0 & 0 \\
     0 & 0 & 0 & 1 & 0 & 0 & 1 & 1 & 0 \\
     0 & 0 & 0 & 0 & 1 & 0 & 1 & 1 & 1 \\
     0 & 0 & 0 & 0 & 0 & 2 & 2 & 2 & 2 \\
     0 & 0 & 0 & 1 & 1 & 2 & 3 & 3 & 3 \\
     0 & 0 & 0 & 1 & 1 & 2 & 3 & 4 & 3 \\
    0 & 0 & 0 & 0 & 1 & 2 & 3 & 3 & 4 \\
};
\draw[black,dashed] 
        (M1-1-1.north west) -| (M1-6-6.south east) -| (M1-1-1.north west);
\end{tikzpicture}
$ = \tilde{C}^{(3)}$ 
\,.
\end{tabular}
\end{equation}
This gives
\begin{equation}\label{eq:A3_third_approx}
C^{(3)} = 
    \begin{bmatrix}
    0 & 0 & 0 & 0 & 0 & 0 \\
    0 & 0 & 0 & 0 & 0 & 0 \\
    0 & 0 & 0 & 0 & 0 & 0  \\
     0 & 0 & 0 & 1 & 0 & 0 \\
     0 & 0 & 0 & 0 & 1 & 0 \\
     0 & 0 & 0 & 0 & 0 & 2\\
    \end{bmatrix},
    \quad
    \begin{aligned}
    x_1^{(3)} = x_1,\quad x_2^{(3)} &= x_2,\quad x_3^{(3)} = x_3,\quad \\ x_4^{(3)} = q^{-1}x_1x_2,\quad x_5^{(3)} &= q^{-1}x_2x_3,\quad x_6^{(3)} = 
    q^{-2}x_1x_2x_3.
    \end{aligned}
\end{equation}
This allows us to write the~cubic approximation (\ref{eq:A3_third_approx}) as
\begin{equation}
\begin{aligned}
\label{eq:PQfactorized}
    P_{Q^{(3)}}(\boldsymbol{x},q) = &\ P_{\text{0-loop}}(x_1,q)\ P_{\text{0-loop}}(x_2,q)\ P_{\text{0-loop}}(x_3,q)\ P_{\text{1-loop}}(q^{-1}x_1x_2,q)\   \\ & \times \ P_{\text{1-loop}}(q^{-1}x_2x_3,q)\ P_{\text{2-loop}}(q^{-2}x_1x_2x_3,q) \\[5pt]
    =
    &\ (x_1;q^2)_{\infty}^{-1}\ (x_2;q^2)_{\infty}^{-1}\ (x_3;q^2)_{\infty}^{-1}\  (x_1x_2;q^2)_{\infty}\  \\ &\times (x_2x_3;q^2)_{\infty}\ [(x_1x_2x_3;q^2)_{\infty}^{-1}\ \times\ (1+O(\boldsymbol{x}^{4}))],
\end{aligned}
\end{equation}
where $(1+O(\boldsymbol{x}^{4}))$ factor comes from infinite $q$-Pochhammers with the~first entry of total degree 4 or higher.
The generating series of corresponding DT invariant reads
\begin{equation}
    \Omega(\boldsymbol{x},q) = -x_1-x_2-x_3+x_1x_2+x_2x_3-x_1x_2x_3 + O(\boldsymbol{x}^{4})\,.
\end{equation}

Summing up, by taking the~union of $n$-th degree approximations we obtain the~infinite diagonal quiver:
\begin{equation}\label{eq:infinite_|Q_0|matrix_A3}
\begin{split}
    C^{(\infty)} &= \diag \big(0,0,0\dashline 1,1\dashline 2 \dashline 3 \dashline 4,4 \dashline 5,5,\dots\big) \\
    \boldsymbol{x}^{(\infty)} &= \big(x_1, x_2, x_3 \dashline q^{-1}x_1x_2, q^{-1}x_2x_3 \dashline
    q^{-2}x_1x_2x_3 \dashline  q^{-3}x_1x_2^2x_3 \dashline \\ 
    &\qquad q^{-4}x_1x_2^2x_3^2, q^{-4}x_1x_2^2x_3^2 \dashline q^{-5}x_1^2x_2^3x_3, q^{-5}x_1x_2^3x_3^2, \dots\big)
\end{split}
\end{equation}

Last but not least, in Tab. \ref{tab:Cinf_A3} we provide a~slightly different presentation of the~infinite diagonal quiver (\ref{eq:infinite_|Q_0|matrix_A3}), where various contributions from one-vertex quivers up to $C^{(\infty)}_{ii}\leq 8$ are segregated.
We believe that this data can be ultimately related to the~poset of generators of the~Lie algebra $\mathfrak{g}_Q$ (see Sec. \ref{sec:gauge}), and leave it for future research.
\begin{table}[h]
    \centering
    \renewcommand{\arraystretch}{1.25}
    \begin{tabular}{||c|c||}
        \hline
        $m$ & unlinkings \\
        \hline
        \hline
        0 & $x_1,\ x_2,\ x_3$ \\
        \hline
        1 & $(x_1,x_2),\ (x_2,x_3)$ \\
        \hline
        2 & $(x_3,(x_1,x_2))$ \\
        \hline
        3 & $((x_1, x_2),(x_2, x_3))$ \\
        \hline
        4 & $((x_1, x_2), (x_3, (x_1, x_2))),\ ((x_2, 
        x_3), (x_3, (x_1, x_2)))$ \\
        \hline
        5 & $((x_1, x_2), ((x_1, x_2), (x_2, x_3))),\ ((x_2, 
        x_3), ((x_1, x_2), (x_2, x_3)))$ \\
        \hline
        6 & $((x_3, (x_1, x_2)), ((x_1, x_2), (x_2, x_3))),
        ((x_1, x_2), ((x_1, x_2), (x_3, (x_1, x_2))))$\\
        & $((x_2, x_3), ((x_1, x_2), (x_3, (x_1, x_2)))),\
        ((x_2, x_3), ((x_2, x_3), (x_3, (x_1, x_2))))$ \\
        \hline
        7 & $((x_1, x_2), ((x_1, x_2), ((x_1, x_2), (x_2, x_3)))),
        ((x_2,x_3), ((x_1, x_2), ((x_1, x_2), (x_2, x_3))))$ \\
        & $((x_2, x_3), ((x_2, x_3), ((x_1, x_2), (x_2, x_3)))),\
        ((x_3, (x_1, x_2)), ((x_1, x_2), (x_3, (x_1, x_2))))$ \\
        & $((x_3, (x_1, x_2)), ((x_2, x_3), (x_3, (x_1, x_2))))$ \\
        \hline
        8 & $((x_3, (x_1, x_2)), ((x_1, x_2), (x_2, x_3))),
        ((x_3, (x_1, x_2)), ((x_1, x_2), ((x_1, x_2), (x_2, x_3))))$ \\
        & $((x_3, (x_1, x_2)), ((x_2, x_3), ((x_1, x_2), (x_2, x_3)))),$\\
        & $(((x_1, x_2), (x_2, x_3)), ((x_1, x_2), (x_3, (x_1, x_2)))),$ \\
        & $(((x_1, x_2), (x_2, x_3)), ((x_2, x_3), (x_3, (x_1, x_2))))$ \\
        \hline
    \end{tabular}
    \caption{$Q^{(\infty)}$ data for symmetric $A_3$ quiver up to $C_{ii}^{\infty}
    \leq 8$. The~rows encode all $C^{(\infty)}_{ii}$-loop quivers contributing to the~infinite factorized form
    (\ref{eq:infinite_factorization}), and the~variable $x$ for each copy of such a~quiver is given by the~bracket expression in the~second column from identification $(x_i,x_j)\simeq q^{-1}x_ix_j$.}
    \label{tab:Cinf_A3}
\end{table}
%

\section{Quivers corresponding to knots}
\label{sec:KQ correspondence}

The knots-quivers correspondence, as its name indicates, expresses various characteristics of knots in terms of those of corresponding symmetric quivers. In particular, it relates the~motivic generating series and motivic DT invariants of a~quiver to the~generating series of HOMFLY-PT polynomials and Labastida-Mari\~{n}o-Ooguri-Vafa (LMOV) invariants of a~knot. The~appearance of symmetric quivers in this context is an~important motivation for our work. Understanding properties of such quivers enables to understand properties of corresponding knots, as well as properties of brane systems in which knots and quivers can be engineered and related to each other.  

The knots-quivers correspondence was discovered in \cite{KRSS1707short,KRSS1707long}. Further developments and elucidations were presented in \cite{PSS1802,EKL1910,EKL1811,JKLNS2105,SW1711,SW2004,EKL2108}. In \cite{PS1811,Kimura:2020qns} the~correspondence was generalized to toric Calabi-Yau manifolds other than the~conifold, and in \cite{Kuch2005,EGGKPSS2110} its version for 3-manifolds that are knot complements was proposed. Diagonalization of quivers discussed in this paper may be of interest in all these contexts.

For a~knot $K\subset S^{3}$, the~HOMFLY-PT polynomial $P_{K}(a,q)$ \cite{HOMFLY,PT} is a~topological invariant which can be calculated via the~skein relation. More generally, colored HOMFLY-PT polynomials $P_{K,R}(a,q)$ are similar polynomial knot invariants that depend also on a~representation $R$ of the~Lie algebra~$\mathfrak{u}(N)$. (In this setting, the~original HOMFLY-PT polynomials correspond to the~fundamental representation.) From the~physical point of view, $P_{K,R}(a,q)$
is the~expectation value of the~knot viewed as a~Wilson line in U$(N)$ Chern-Simons gauge theory \cite{Witten_Jones}. The~HOMFLY-PT generating series is given by
\begin{equation}
P_{K}(\lambda,a,q)=\sum_{r=0}^{\infty}P_{K,r}(a,q)\lambda^{-r},
\end{equation}
where $P_{K,r}(a,q)$ are HOMFLY-PT polynomials colored by the~totally
symmetric representations $S^{r}$. The~unusual expansion variable with the~negative power comes from the~necessity of resolving the~clash of different conventions present in the~literature and avoiding the~confusion with the~quiver variables. For more detailed discussion of all conventions see \cite[Sec. 2.1]{Kuch2005} and references therein.

The LMOV invariants $N_{r,i,j}$ \cite{OV9912,LM01,LMV00} are numbers that give the~following expression for the~HOMFLY-PT generating
series\footnote{Note that in this paper the~product form with quantum dilogarithms serves as a~basis. In consequence, we have $\sum_{j\in\mathbb{Z}}N_{r,i,j}q^j=\sum_{k\in\mathbb{Z}}N'_{r,i,k}q^{k+1}$, where $N'_{r,i,k}$ is the~LMOV invariant in the~notation from \cite{KRSS1707long}.}:
\begin{equation}\label{eq:P_K=00003DExp}
P_{K}(\lambda,a,q)=\prod_{r\geq 1}\prod_{i,j\in \mathbb{Z}}(\lambda^{-r} a^i q^j;q^2)_{\infty}^{N_{r,i,j}}
\end{equation}
We can assemble $N_{r,i,j}$  into the~LMOV generating series $N(\lambda,a,q)=\sum_{r,i,j}N_{r,i,j}\lambda^{-r}a^{i}q^{j}$.
From the~physical point of view, LMOV invariants count BPS
states in the~effective
3d $\mathcal{N}=2$ theory on the~world-volume of M5-brane wrapped
on the~knot conormal inside the~resolved conifold \cite{OV9912}.

The~knots-quivers correspondence\,\cite{KRSS1707short,KRSS1707long}
is a~conjecture that for each knot~$K$ there exist a~quiver~$Q$
and integers $n_{i}$, $a_{i}$, $l_{i}$, $\ensuremath{i\in Q_{0}}$,
such that
\begin{equation}
P_{K}(\lambda,a,q)=\left.P_{Q}(\boldsymbol{x},q)\right|_{x_{i}=\lambda^{n_{i}}a^{a_{i}}q^{l_{i}}}.
\end{equation}
If we substitute (\ref{eq:P_K=00003DExp}) and (\ref{PQ-product}),
we obtain the\,knots-quivers correspondence at the~level of LMOV
and DT invariants:
\begin{equation}
N(\lambda,a,q)=\left.\Omega(\boldsymbol{x},q)\right|_{x_{i}=\lambda^{n_{i}}a^{a_{i}}q^{l_{i}}}.
\end{equation}
Since DT invariants are integer, this equation implies integrality
of $N_{r,i,j}$ -- this means that the~knots-quivers correspondence automatically proves the~LMOV conjecture.

In the~rest of this section we apply the~framework of infinite diagonal quivers and $m$-loop quivers to the~knots-quivers correspondence.
Quivers corresponding to knots are in general much more complicated than examples from the~previous section. In consequence, we will focus on the~reduced normalization\footnote{For the~discussion of the~normalizations of HOMFLY-PT polynomials see \cite[Sec. 4.5]{KRSS1707long}. The~fact that we use the~reduced normalization implies that the~BPS spectrum we consider is a~subset of the~BPS spectrum of the~unreduced case built from the~half of the~generators. We do it to avoid very tedious computations, but all our methods can be applied also in the~unreduced normalization.} and simplest nontrivial examples, namely the~trefoil and figure-eight. Moreover, even in these cases the~matrix $\tilde{C}^{(2)}$ exceeds $50\times50$ entries. Therefore, we will compute $C^{(2)}$ and, if possible, $C^{(3)}$ in the~most efficient way, (un)liking only those non-diagonal entries that are necessary.

\subsection{Trefoil knot}

In this case the~adjacency matrix and the~change of variables are given by \cite{KRSS1707long}
\begin{equation}\label{eq:trefoil_quiver_{|Q_0|}atrix}
    C=\left[
\begin{array}{ccc}
 0 & 1 & 1 \\
 1 & 2 & 2 \\
 1 & 2 & 3 \\
\end{array}
\right]~,
\qquad
\left[
\begin{array}{c}
x_1 \\
x_2 \\
x_3 \\
\end{array} \right]  = \left[
\begin{array}{c}
\lambda^{-1} a^2 q^{-2}\\
\lambda^{-1} a^2 \\
\lambda^{-1} a^4 q^{-3} \\
\end{array} \right].
\end{equation}

The first degree approximation is immediate, as it coincides with the~diagonal of the~initial quiver (\ref{eq:trefoil_quiver_{|Q_0|}atrix}) with $\boldsymbol{x}^{(1)} = (x_1,x_2,x_3)$. The~second order approximaton comes from unlinking all arrows in (\ref{eq:trefoil_quiver_{|Q_0|}atrix}), which gives
\begin{equation}
\tilde{C}^{(1)} = 
\begin{tikzpicture}[
    node distance=1mm and 0mm,
    baseline]
\matrix (M1) [matrix of nodes,{left delimiter=[},{right delimiter=]}]
{
0 & 0 & 0 & 0 & 0 & 0 & 0 \\
 0 & 2 & 0 & 2 & 2 & 3 & 2 \\
 0 & 0 & 3 & 3 & 3 & 4 & 3 \\
 0 & 2 & 3 & 3 & 3 & 5 & 5 \\
 0 & 2 & 3 & 3 & 4 & 5 & 5 \\
 0 & 3 & 4 & 5 & 5 & 8 & 7 \\
 0 & 2 & 3 & 5 & 5 & 7 & 6 \\
};
\draw[black,dashed] 
        (M1-1-1.north west) -| (M1-3-3.south east) -| (M1-1-1.north west);
\end{tikzpicture}~.
\end{equation}
Since the~seven diagonal entries of $\tilde{C}^{(1)}$ are the~only ones which contribute to quadratic terms $x_ix_j$ in $P_Q(\boldsymbol{x},q)$, we do not have to write the~whole $\tilde{C}^{(2)}$, but we can conclude that
\begin{equation}
\begin{split}
    C^{(2)} &= \diag \big(0, 2, 3 \dashline 3, 4, 8, 6\big) \\
    \boldsymbol{x}^{(2)} &= \left(x_1,x_2,x_3 \dashline q^{-1}x_1x_2,q^{-1}x_1x_3,q^{-1}x_2x_3,q^{-1}x_2x_3\right).
\end{split}
\end{equation}

Following  Sec.~\ref{sec:Integrality} and expanding the~product of $C_{ii}$-loop quivers, we obtain
\begin{align}
    \Omega(\boldsymbol{x},q) = & -x_1 - q^2x_2  + q^3x_3 \\
    & + q^2x_1x_2  - q^3x_1x_3 + q^4x_2^2 - q^5x_2x_3  - q^7x_2x_3 + q^8x_3^2 + O\left(\boldsymbol{x}^3\right)~. \nonumber
\end{align}
The application of the~change of variables \eqref{eq:trefoil_quiver_{|Q_0|}atrix} translates DT invariants into LMOV invariants:
\[
    N(\lambda, a, q) =  -a^2 q^{-2} \left(1+a^2 q^5+q^6\right) \lambda^{-1} + a^4 q \left(a^2+q\right) \left(1+q^6+a^2 q^7\right) \lambda^{-2} + O\left(\lambda^{-3}\right)~.
\]

In order to find the~third degree approximation, we have to unlink all arrows between the~four new nodes corresponding to diagonal entries $(3,4,8,6)$ and the~initial triple corresponding to $(0,2,3)$. For clarity, we highlight the~unlinked non-diagonal entries in red: 
\begin{equation}
\tilde{C}^{(1)} =
\begin{tikzpicture}[
    node distance=1mm and 0mm,
    baseline]
\matrix (M1) [matrix of nodes,{left delimiter=[},{right delimiter=]}]
{
0 & 0 & 0 & 0 & 0 & 0 & 0 \\
 0 & 2 & 0 & \textcolor{red}{2} & \textcolor{red}{2} & \textcolor{red}{3} & \textcolor{red}{2} \\
 0 & 0 & 3 & \textcolor{red}{3} & \textcolor{red}{3} & \textcolor{red}{4} & \textcolor{red}{3} \\
 0 & \textcolor{red}{2} & \textcolor{red}{3} & 3 & 3 & 5 & 5 \\
 0 & \textcolor{red}{2} & \textcolor{red}{3} & 3 & 4 & 5 & 5 \\
 0 & \textcolor{red}{3} & \textcolor{red}{4} & 5 & 5 & 8 & 7 \\
 0 & \textcolor{red}{2} & \textcolor{red}{3} & 5 & 5 & 7 & 6 \\
};
\draw[black,dashed] 
        (M1-1-1.north west) -| (M1-3-3.south east) -| (M1-1-1.north west);
\end{tikzpicture}~.
\end{equation}
This procedure results in $29\times 29$ matrix which is too large to write here. However, its main diagonal and the~corresponding change of variables gives us the~third order \hfill approximation:
\begin{align}
    C^{(3)} = & \diag \big(0, 2, 3 \dashline 3, 4, 8, 6 \dashline \nonumber\\
    & \qquad \qquad 8, 6, 9, 7, 15, 13, 11, 11, 9, 11, 9, 7, 12, 10, 8, 18, 16, 14, 12, 14, 12, 10  \big), \\
    \boldsymbol{x}^{(3)} = & \big( x_1,x_2,x_3 \dashline q^{-1}x_1x_2,q^{-1}x_1x_3,\{q^{-1}x_2x_3\}_2 \dashline \nonumber\\
    & \quad \{q^{-2}x_1x_2^2\}_2,\{q^{-2}x_1x_2x_3\}_2,\{q^{-2}x_2^2x_3\}_3,\{q^{-2}x_2^2x_3\}_2,\{q^{-2}x_1x_2x_3\}_3,\{q^{-2}x_1x_3^2\}_3,\nonumber\\
    &\{q^{-2}x_2x_3^2\}_4,\{q^{-2}x_2x_3^2\}_3  \big),\nonumber
\end{align}
where we denote $\{\alpha\}_n=\overbrace{\alpha,\dots,\alpha}^{\text{$n$ times}}$ for any monomial $\alpha$ .

In turn, using Theorem \ref{thm:KS} we compute
\begin{equation}
    \begin{aligned}
        \Omega(\boldsymbol{x},q) = & \ 
        -x_1-q^2 x_2+q^3 x_3 \\
        & +q^2 x_1 x_2-q^3 x_1 x_3+q^4 x_2^2-q^5 x_2 x_3+q^8 x_3^2 \\
        & -(q^4+q^6) x_1 x_2^2 - (q^6+q^8)x_1 x_3^2 + (2q^5+2q^7+q^9)x_1 x_2 x_3 + O\left(\boldsymbol{x}^4\right)
    \end{aligned}
\end{equation}
The generating series of the~corresponding LMOV invariants reads
\begin{equation}
    \begin{aligned}
        N(\lambda,a, q) = & -a^2 q^{-2} \left(1+a^2 q^5+q^6\right) \lambda^{-1} + a^4 q \left(a^2+q\right) \left(1+q^6+a^2 q^7\right) \lambda^{-2} \\
         & - a^6 q^{11} \left(a^6+a^6 q^2+q^3+a^6 q^6\right) \lambda^{-3} + O\left(\lambda^{-4}\right)\,.
    \end{aligned}
\end{equation}

\subsection{Figure-eight knot}

This is the~first example in our selection where we encounter both ordinary and negative arrows in $Q$.
This indicates necessity to use both unlinking and linking in order to compute~$Q^{(n)}$,
following the~rules of diagonalization given in Def. \ref{def:Rules of diagonalization}.
The adjacency matrix and the~change of variables are given by
\begin{equation}
\label{eq:figure eight matrix}
    C=\left[
\begin{array}{ccccc}
 0 & -1 & -1 & 0 & 0 \\
 -1 & -2 & -2 & -1 & 0 \\
 -1 & -2 & -1 & -1 & 0 \\
 0 & -1 & -1 & 1 & 1 \\
 0 & 0 & 0 & 1 & 2 \\
\end{array}
\right]~,
\qquad
\left[
\begin{array}{c}
x_1 \\
x_2 \\
x_3 \\
x_4 \\
x_5 \\
\end{array} \right] = \left[
\begin{array}{c}
\lambda^{-1} \\
\lambda^{-1} a^{-2} q^2 \\
\lambda^{-1} q^{-1} \\
\lambda^{-1} q \\
\lambda^{-1} a^2 q^{-2} \\
\end{array} \right]
.
\end{equation}
The first degree approximation is immediately given by $C^{(1)} = \diag(0,-2,-1,1,2)$, $\boldsymbol{x}^{(1)} = (x_1,\dots,x_5)$. $C^{(1)}$ is a~diagonal submatrix of
\begin{equation}
\tilde{C}^{(1)} = 
\begin{tikzpicture}[
    node distance=1mm and 0mm,
    baseline]
\matrix (M1) [matrix of nodes,{left delimiter=[},{right delimiter=]}]
{
0 & 0 & 0 & 0 & 0 & $-$1 & $-$1 & 0 & 0 & 0 & 0 & 0 \\
 0 & $-$2 & 0 & 0 & 0 & $-$3 & $-$2 & $-$4 & $-$3 & $-$3 & $-$1 & 0 \\
 0 & 0 & $-$1 & 0 & 0 & $-$3 & $-$2 & $-$3 & $-$2 & $-$2 & $-$2 & 0 \\
 0 & 0 & 0 & 1 & 0 & $-$1 & $-$1 & $-$2 & 0 & 0 & 0 & 1 \\
 0 & 0 & 0 & 0 & 2 & 0 & 0 & 0 & 0 & 1 & 1 & 2 \\
 $-$1 & $-$3 & $-$3 & $-$1 & 0 & $-$4 & $-$4 & $-$6 & $-$6 & $-$4 & $-$4 & $-$1 \\
 $-$1 & $-$2 & $-$2 & $-$1 & 0 & $-$4 & $-$3 & $-$4 & $-$4 & $-$3 & $-$3 & $-$1 \\
 0 & $-$4 & $-$3 & $-$2 & 0 & $-$6 & $-$4 & $-$7 & $-$7 & $-$6 & $-$5 & $-$2 \\
 0 & $-$3 & $-$2 & 0 & 0 & $-$6 & $-$4 & $-$7 & $-$5 & $-$5 & $-$3 & 0 \\
 0 & $-$3 & $-$2 & 0 & 1 & $-$4 & $-$3 & $-$6 & $-$5 & $-$3 & $-$2 & 1 \\
 0 & $-$1 & $-$2 & 0 & 1 & $-$4 & $-$3 & $-$5 & $-$3 & $-$2 & $-$2 & 1 \\
 0 & 0 & 0 & 1 & 2 & $-$1 & $-$1 & $-$2 & 0 & 1 & 1 & 4 \\
};
\draw[black,dashed] 
(M1-1-1.north west) -| (M1-5-5.south east) -| (M1-1-1.north west);
\end{tikzpicture}
\ ,
\end{equation}
obtained from $C$ by the~successive application of (un)linking. The~diagonal of $\tilde{C}^{(1)}$ together with the~change of variables coming from (un)linking gives us the~second degree approximation (the matrix $\tilde{C}^{(2)}$ is of size $125 \times 125$):
\begin{equation}
\begin{split}
    C^{(2)} &= \diag\big(0, -2, -1, 1, 2 \dashline -4, -3, -7, -5, -3, -2, 4\big)\, .   \\
        \boldsymbol{x}^{(2)} & = \big( x_1, x_2, x_3, x_4, x_5\dashline x_1x_2, x_1x_3 , \{x_2x_3\}_2 , x_2x_4 ,x_3x_4 ,q^{-1}x_4x_5\big). 
\end{split}
\end{equation}
%
%
This comes from the~fact that all nodes whose leading term is of the~form $x_ix_j$, $i\neq j$, come from
(un)linking of the~nodes of $Q$.
Using this data and Theorem \ref{thm:KS}, we can write the~generating seires of DT invariants up to quadratic terms:
\begin{equation}
\begin{aligned}
\Omega(\boldsymbol{x},q) = &\ -x_1 - q^{-2}x_2 + q^{-1}x_3 + qx_4 - q^2x_5 \\ 
 & - q^{-4}x_1x_2 + q^{-3}x_1x_3 - q^{-8}x_2^2 + (q^{-7} + q^{-5})\,x_2x_3 + q^{-3}x_2x_4 - q^{-4}x_3^2 - q^{-2}x_3x_4 \\
 & - q^3x_4x_5 + q^4x_5^2 + O\left(\boldsymbol{x}^3\right)\,.
\end{aligned}
\end{equation}
As a~consistency check, note that terms of the~form $x_ix_j$ with $i\neq j$ precisely correspond to seven extra
nodes in $C^{(2)}$. 
The application of the~change of variables \eqref{eq:figure eight matrix} gives the~following LMOV invariants:
\begin{equation}
    \begin{aligned}
       N(\lambda^{-1},a,q) = & - a^{-2} q^{-3}\left(a^2+a^4 q+a^2 q^3+q^5+a^2 q^6\right) \lambda^{-1} \\
        & - a^{-4} q^{-9} \left(1+a^2 q\right) \left(a^6+a^4 q+a^4 q^4+a^4 q^7-q^{13}\right) \lambda^{-2} + O\left(\lambda^{-3}\right)\,.
    \end{aligned}
\end{equation}
%

\section{Quivers corresponding to knot complements}
\label{sec:FK quivers}

$F_K$~invariants were introduced in~\cite{GM1904} as knot complement versions of $\hat{Z}$~invariants \cite{GPV1602,GPPV1701}:
\begin{equation}
F_{K}=\hat{Z}\left(S^{3}\backslash K\right).
\end{equation} 
From the~physical point of
view, the~$F_K$~invariant arises from the~reduction of 6d $\mathcal{N}=(0,2)$
theory describing M5-branes on the~3-manifold with the~topology of
the~knot complement \cite{GM1904}. The~$a$-deformed $F_K$ invariants were introduced in \cite{GGKPS20xx}. In \cite{Kuch2005} the~knots-quivers correspondence was generalized to knot complements using the~$a$-deformed version of~$F_{K}$ (which for simplicity we will call just $F_K$ invariant). More specifically, the~author conjectured -- and proved for the~complements of $(2,2p+1)$ torus knots -- that after appropriate change of variables the~$F_K$ invariant can be expressed as a~motivic generating series of some quiver $Q$: 
\begin{equation}
\label{eq:FK-Q correspondence}
F_{K}(\mu,a,q)=
\sum_{d_1,\dots,d_{|Q_0|}\geq 0}(-q)^{\sum_{i,j=1}^{|Q_0|} C_{ij}d_i d_j}\prod_{i=1}^{|Q_0|}\frac{\mu^{n_{i}d_{i}}a^{a_{i}d_{i}}q^{l_{i}d_{i}}}{(q^{2};q^{2})_{d_i}}
=\left.P_{Q}(\boldsymbol{x},q)\right|_{x_{i}=\mu^{n_{i}}a^{a_{i}}q^{l_{i}}}.
\end{equation}
As a~consequence, the~knot complement analogues of LMOV invariants $N(\mu,a,q)$ were introduced basing on DT invariants of corresponding quivers: 
\begin{equation}
\label{eq:FK-Q for BPS invariants}
N(\mu,a,q)=\sum_{r,i,j}N_{r,i,j}\mu^{r}a^{i}q^{j}=\left.\Omega(\boldsymbol{x},q)\right|_{x_{i}=\mu^{n_{i}}a^{a_{i}}q^{l_{i}}}.
\end{equation}
Further developments, including explicit formulas for many knot complements and generalisation to different branches of $A$-polynomials were presented in \cite{EGGKPSS2110}.

In the~rest of this section we apply the~framework of infinite diagonal quivers and $m$-loop quivers to the~simplest nontrivial cases of quivers corresponding to the~trefoil and figure-eight knot complements.\footnote{More precisely, we consider quivers corresponding to the $F_K$ invariants associated to the abelian branches of the $A$-polynomials for these knots. For more details see \cite{EGGKPSS2110}.}

\subsection{Trefoil knot complement}

In case of the~trefoil knot complement, $F_{3_1}$ can be written in the~form \eqref{eq:FK-Q correspondence} for the~following quiver and the~change of variables:\footnote{The difference between the~change of variables \eqref{eq:trefoil knot complement} and \cite{Kuch2005} comes from the~conventional switch $q\leftrightarrow q^2$ and $a\leftrightarrow a^2$.}\cite{Kuch2005}
\begin{equation}
\label{eq:trefoil knot complement}
    C=\left[
\begin{array}{cccc}
 0 & -1 & 0 & -1 \\
 -1 & -1 & 0 & -1 \\
 0 & 0 & 1 & 0 \\
 -1 & -1 & 0 & 0 \\
\end{array}
\right],
\qquad
\left[
\begin{array}{c}
x_1 \\
x_2 \\
x_3 \\
x_4 \\
\end{array} \right] = \left[
\begin{array}{c}
\mu^2 a^2 \\
\mu^2 q^{3} \\
\mu a^2 q^{-1} \\
\mu q^2 \\
\end{array} \right].
\end{equation}
In this example we explicitly present two steps of the~calculation -- the~second and third degree approximations, which are computable in reasonable time.
After linking all non-diagonal entries, we obtain the~second degree approximation:
\begin{equation}
    \begin{aligned}
        C^{(2)}=&\ \diag(0, -1, 1, 0 \dashline -3, -2, -3)\,, \\
         \boldsymbol{x}^{(2)}=&\ (x_1, x_2, x_3, x_4 \dashline x_1 x_2, x_1 x_4, x_2 x_4)\,.
    \end{aligned}
\end{equation}
Following the~procedure of expanding the~product of $C_{ii}$-loop quivers, described in Sec.~\ref{sec:Integrality}, we get
\begin{equation}\label{eq:trefoil complement DT quadratic}
\begin{aligned}
    \Omega(\boldsymbol{x},q) = &\ -x_1 + q^{-1} x_2 + q x_3 - x_4 
    \\
    &\ + q^{-3} x_1 x_2 - q^{-4} x_2^2 - q^{-2} x_1 x_4  + q^{-3} x_2 x_4 + O\left(\boldsymbol{x}^3\right)\,.
\end{aligned}
\end{equation}
Combining the~above result with the~definition \eqref{eq:FK-Q for BPS invariants} and change of variables \eqref{eq:trefoil knot complement}, we obtain the~trefoil knot complement analogues of LMOV invariants:
\begin{equation}
    N(\mu, a, q) =  (a^2-q^2)\, \mu-(a^2-q^2)\, \mu^2 + O(\mu^3)~.
\end{equation}
Following analogous steps, we obtain the~third degree approximation:
\begin{equation}
    \begin{aligned}
       C^{(3)}=&\ \diag(0, -1, 1, 0 \dashline -3, -2, -3 \dashline -5, -4, -8, -6, -5, -8, -6, -7, -5, -4, -5)\,, \\
       \boldsymbol{x}^{(3)}=&\ (x_1, x_2, x_3, x_4 \dashline x_1 x_2, x_1 x_4, x_2 x_4 \dashline x_1^2 x_2,x_1^2 x_4,\{x_1 x_2^2\}_2,x_1 x_2 x_4,\{x_2^2 x_4\}_2, \\
      &\quad \{x_1 x_2 x_4\}_2,x_1 x_4^2,x_2 x_4^2)\,.
    \end{aligned}
\end{equation}
This allows to refine the~expression for the~generating series given in (\ref{eq:trefoil complement DT quadratic}):
\begin{equation}
    \begin{aligned}
        \Omega(\boldsymbol{x},q) = &
        - x_1 + q^{-1} x_2 + q x_3 - x_4 \\
        & + q^{-3} x_1 x_2 - q^{-2} x_1 x_4 - q^{-4} x_2^2+q^{-3} x_2 x_4 \\
        & + q^{-5} x_1^2 x_2 - q^{-4} x_1^2 x_4 - (q^{-8} + q^{-6})\, x_1 x_2^2 + (q^{-7}+2 q^{-5})\, x_1 x_2 x_4 \\
        & -q^{-4} x_1 x_4^2+q^{-9} x_2^3 - (q^{-8} + q^{-6})\, x_2^2 x_4+q^{-5} x_2 x_4^2
         +O(\boldsymbol{x}^4)~.
    \end{aligned}
\end{equation}
After specialization of $x_i$, given by (\ref{eq:trefoil knot complement}), they take form
\begin{equation}
    N( \mu,a, q) = \left(a^2-q^2\right) \mu - \left(a^2 - q^2\right) \mu^2 - \left(a^2 - q^2\right) \mu^3 + O(\mu^4)~.
\end{equation}
The presence of $(a^2 - q^2)$ factor in the~expression for $N(\mu, a, q)$ is a~trace of trivialization $F_K(\mu,a=q,q)=1$ conjectured in \cite{GGKPS20xx} for any knot complement. However, the~lack of other dependence on $a$ and $q$ suggests that the~analogous of LMOV invariants for the~complement of $3_1$ satisfy the~following property
\begin{equation}\label{eq:Recip LMOV 3_1}
    N_{S^3 \setminus 3_1}(\mu, a^{-1}, q^{-1}) = (-a^{-2} q^{-2} )\, N_{S^3 \setminus 3_1}(\mu, a, q).
\end{equation}
%

\subsection{Figure-eight knot complement}

For the~figure-eight knot complement, one can write $F_{4_1}$ in the~form \eqref{eq:FK-Q correspondence} using the~following quiver and the~change of variables:\footnote{The difference between the~change of variables \eqref{eq:Quiver41KC} and \cite{EGGKPSS2110} comes from the~conventional switch $q\leftrightarrow q^2$ and $a\leftrightarrow a^2$.}  \cite{EGGKPSS2110}

\begin{equation}\label{eq:Quiver41KC}
    C=\left[
\begin{array}{cccccc}
 0 & 0 & 0 & 1 & 0 & 0 \\
 0 & 0 & -1 & -1 & 0 & 0 \\
 0 & -1 & 0 & 0 & 0 & 0 \\
 1 & -1 & 0 & 1 & 1 & 0 \\
 0 & 0 & 0 & 1 & 1 & 0 \\
 0 & 0 & 0 & 0 & 0 & 1 \\
\end{array}
\right],
\qquad
\left[
\begin{array}{c}
x_1 \\
x_2 \\
x_3 \\
x_4 \\
x_5 \\
x_6 \\
\end{array} \right] = \left[
\begin{array}{c}
\mu q^2 \\
\mu q^2 \\
\mu q^2 \\
\mu a^2 q^{-1} \\
\mu a^2 q^{-1} \\
\mu a^2 q^{-1} \\
\end{array} \right].
\end{equation}
The second degree approximation obtained after (un)linking all non-diagonal entries reads
\begin{equation}
\begin{split}
    C^{(2)} &= {\rm diag}(0, 0, 0, 1, 1, 1\dashline 2, -2, -1, 3)~,\\
    \boldsymbol{x}^{(2)}&=(x_1, x_2, x_3, x_4, x_5, x_6 \dashline q^{-1} x_1 x_4, x_2 x_3, x_2 x_4, q^{-1} x_4 x_5).
\end{split}
\end{equation}
After the~application of Theorem \ref{thm:KS}, this leads to
\begin{align}\label{eq:DT41KC2ndOrder}
    \Omega(\boldsymbol{x},q) = & - x_1 - x_2 - x_3 + q x_4 + q x_5 + q x_6 \\
     & -q x_1 x_4-q^{-2} x_2 x_3+q^{-1} x_2 x_4+q^2 x_4 x_5 + O\left(\boldsymbol{x}^3\right)~. \nonumber
\end{align}
If we put this equation together with \eqref{eq:trefoil knot complement} into
\eqref{eq:FK-Q for BPS invariants}, we obtain 
\begin{equation}
    N(\mu, a, q) = 3 \left(a^2 - q^2\right) \mu + \left(1+a^2\right) \left(a^2 - q^2\right) \mu ^2 + O\left(\mu^3\right)~.
\end{equation}
Following similar steps, we get the third degree approximation:
\begin{equation}
\begin{split}
    C^{(3)} &= {\rm diag}(0, 0, 0, 1, 1, 1\dashline 2, -2, -1, 3 \dashline 0, -4, -3, -4, 4, -3, 5, 4, 1, 5)~,\\
    \boldsymbol{x}^{(3)}&=(x_1, x_2, x_3, x_4, x_5, x_6 \dashline q^{-1} x_1 x_4, x_2 x_3, x_2 x_4, q^{-1} x_4 x_5 \dashline q^{-1} x_1 x_2 x_4, x_2^2 x_3, x_2^2 x_4,\\
    & \qquad x_2 x_3^2, q^{-2} x_1 x_4^2, x_2 x_3 x_4, q^{-2} x_4^2 x_5, q^{-2} x_1 x_4 x_5, q^{-1} x_2 x_4 x_5, q^{-2} x_4 x_5^2).
\end{split}
\end{equation}
This allows us to compute more DT invariants, extending \eqref{eq:DT41KC2ndOrder} to
\begin{equation}
\begin{aligned}
    \Omega(\boldsymbol{x},q) = & -x_1-x_2-x_3+q x_4+q x_5+q x_6 \\
      & -q x_1 x_4-q^{-2} x_2 x_3+q^{-1} x_2 x_4+q^2 x_4 x_5 \nonumber\\
      & +q^2 x_4 x_5-q^{-1} x_1 x_2 x_4-q^2 x_1 x_4^2-q^2 x_1 x_4 x_5-q^{-4} x_2^2 x_3+q^{-3} x_2^2 x_4-q^{-4} x_2 x_3^2 \nonumber\\
      & +q^{-3} x_2 x_3 x_4+x_2 x_4 x_5+q^3 x_4^2 x_5+q^3 x_4 x_5^2+O\left(\boldsymbol{x}^4\right)~. \nonumber
\end{aligned}
\end{equation}
    Finally, we perform the~change of variables indicated in \eqref{eq:Quiver41KC} to derive the~analogue of LMOV invariants of $F_{4_1}$:
\begin{align}
    N(\mu, a, q) = & 3 \left(a^2 - q^2\right) \mu + \left(1+a^2\right) \left(a^2 - q^2\right) \mu ^2 \\
     & + \left(2+a^2+2 a^4\right) \left(a^2 - q^2\right) \mu ^3 + O\left(\mu^4\right)~.\nonumber
\end{align}

Similarly to the~LMOV invariants for the~complement of $3_1$  (which are reciprocal polynomials, see Eq. \eqref{eq:Recip LMOV 3_1}), we conjecture that the~above expression enjoys the~property
\begin{equation}\label{eq:Recip LMOV 4_1}
    N_{S^3 \backslash 4_1}\left(\mu, a^{-1}, q^{-1}\right) = \left(-q^{-2}\right) N_{S^3 \backslash 4_1}\left(a^2 \mu, a, q\right).
\end{equation}


\section*{Acknowledgments}

This work has been supported by the~TEAM programme of the~Foundation for Polish Science co-financed by the~European Union under the~European Regional Development Fund (POIR.04.04.00-00-5C55/17-00), Vidi grant (number 016.Vidi.189.182) from the
Dutch Research Council (NWO), and Excellence Initiative: Research University programme financed by the Polish Ministry of Education and Science (01/IDUB/2019/94). P.S. also acknowledges the~support of Fulbright STEM Impact Award and hospitality of the High Energy Theory Group at Harvard University.


\appendix


\section{Combinatorial model for \texorpdfstring{$(-m)$}{-m}-loop quivers}
\label{app:combinatorial}

In Sec. \ref{subsec:m-loop} we proved  Lemma~\ref{lma:mloop} using a~simple $q$-Pochhammer identity. However, we can also show how the~relation between DT invariants for $(-m)$- and $(m+1)$-loop quivers is realized at the~combinatorial level. The~base for this construction is a~combinatorial model for $m$-loop quivers introduced in \cite{Rei12}. In \cite{KS1608} it was connected to the~quantum version of extremal $A$-polynomials and extremal LMOV invariants of knots \cite{GKS1504}. The~quantum $A$-polynomials annihilate HOMFLY-PT polynomials \cite{Gar0306,Guk0306,AV1204,GLL1604} and have quiver analogues that annihilate motivic generating series \cite{EKL1910,Larraguivel:2020sxk,Noshchenko:2020lfq}. 

The reasoning of the~proof below can be summarized in the~following way.
The relation between $m$-loop quiver model \cite{Rei12} and the~model associated to extremal quantum $A$-polynomial with $m$'th $q$-power \cite{KS1608} was described in \cite[App. B]{KS1608} for $m\geq 0$. However, extremal quantum $A$-polynomials can be considered also with a~$q$-power of $(-m)$ (see Eq.~\eqref{eq:Apoly}). Then, following the~steps of \cite[App. B]{KS1608}, we can relate the~case of $(-m)$ to the~one corresponding to $m+1$.

\begin{proof}[Combinatorial proof of Lemma~\ref{lma:mloop}]
In order to follow \cite[App. B]{KS1608} and the~conventions used in that paper, we rescale our variable $x$ by $q^{m+1}$ and sum over $r$ instead of $d$:
\begin{equation}
P_{\text{$(-m)$-loop}}(x' q^{m+1},q)= 
\sum_{r=0}^{\infty} \frac{(-q)^{(-m)r^2}q^{(m+1)r}}{(q^2;q^2)_r}\,(x')^r\,.
\end{equation}
Then, we define
\begin{equation}
   Y(x',q) =\frac{P_{\text{$(-m)$-loop}}(x' q^{m+3},q)}{P_{\text{$(-m)$-loop}}(x' q^{m+1},q)}~,
\end{equation}
which satisfies the~following $A$-polynomial equation:
\begin{equation}
    1-Y(x',q)+(-1)^{m+1}qx'
    \prod_{i=1}^{m}Y(q^{-2i}x',q)^{-1}=0~.
    \label{eq:Apoly}
\end{equation}
Using the~series expansion, $Y(x',q)=\sum_{n=0}^\infty Y_n(q)(x')^n~,$
Eq.~(\ref{eq:Apoly}) implies that
\begin{equation}
    Y_n(q) = -\sum_{\substack{a+b=n\\ 0<ab}} Y_a(q)\sum_{k_1+\cdots+k_{m}=b}
    \prod_{i=1}^{m}q^{-2ik_i}Y_{k_i}(q)~,
    \label{eq:recursion}
\end{equation}
with $Y_0(q)=1$ and $Y_1(q)=(-1)^{m+1}q$.\footnote{Note that on the~right-hand side there are $m+1$ factors of $Y$, whereas in \cite[Eq. (4.13)]{KS1608} there are $m$ of them. This is a~structural manifestation of the~Lemma \ref{lma:mloop}.}
For example, for $n=2$ we have only one option $a=1$ and $b=1$, whence
\begin{equation}
    Y_2(q)=-Y_1(q)\sum_{i=1}^{m}q^{-2i}Y_1(q)=-Y_1(q)^2\sum_{i=1}^{m}q^{-2i}=-q^2\sum_{i=1}^{m}q^{-2i}~.
\end{equation}

Using Eq.~(\ref{eq:recursion}), analogously to the~construction from Ref.~\cite{KS1608}, we can represent $Y_n(q)$ as a~signed list $\phi=(-1)^{mn+1}[\phi_1,\cdots, \phi_{n}]$.
In order to do it, we start from
\begin{equation}
    T_0 = \{[]\}~, \qquad T_1 = \{(-1)^{m+1}[1]\}~,
\end{equation}
and define sets $T_2, T_3,\cdots$ recursively by
\begin{eqnarray}
\label{eq:Tn}
    T_n&=&\Bigg\{ - \phi_{(a)}*\left( \phi_{(k_{m})}-2m\right)*\cdots*\left(\phi_{(k_1)}-2 \right)\\
    \nonumber
    &|& \forall\left( a+b=n  \wedge 0<ab\right)  \wedge \left(\forall k_1+\cdots+k_{m}=b\wedge \phi_{(k_i)}\in T_{k_i},~  i=1,\cdots,m\right) \Bigg\}~,
\end{eqnarray}
where $*$ is a~concatenation of lists:
\begin{equation}
    \phi_{(1)} * \phi_{(2)} = [\phi_{(1)},\phi_{(2)}] = [\phi_{(1),1},\dots,\phi_{(1),n_1},\phi_{(2),1},\dots,\phi_{(2),n_2}]~,
\end{equation}
and subtracting a~number from a~list means a~subtraction for each entry.
All powers in Eq.~(\ref{eq:recursion}) sum up to $n$ and eventually all terms are expressed as powers of $Y_1(q)=(-1)^{m+1}q$. 
Since each $Y_*$ term contributes a~$(-1)$, 
then the~sign of a~list of length $n$ is $(-1)^{n(m+1)+n+1}=(-1)^{1+(2+m) n}=(-1)^{1+mn}$.
The weight of a~list is 
${\rm wt}(\phi)=\sum_{i=1}^{n} \phi_i$, so that
\begin{equation}
    Y_n(q)=\sum_{\phi\in T_n}{\rm sgn}(\phi) q^{{\rm wt}(\phi)}~.
\end{equation}
Note that Eq.~(\ref{eq:Tn}) implies that for $n>0$ we have
\begin{equation}
    \phi_0=1~, \qquad \phi_i\leq\phi_0 \qquad {\rm for} \qquad i=1,\cdots,n-1, \qquad |\phi_{i+1}-\phi_i|\leq 2m~.
\end{equation}

Now we are ready to define a~map $\varphi$ from 
$(-m)$-loop case to $(m+1)$-loop case: 
\begin{equation}
    \varphi(\phi)=-[2-\phi_1,2-\phi_2,\cdots,2-\phi_{n}]~.
\end{equation}
Comparing Eq.~\eqref{eq:Tn} with the~rules of creating primary lists\footnote{We say that the~list is primary if it cannot be expressed as a~concatenation of smaller lists.} given in \cite[Eq. (4.18)]{KS1608}, one can check that $\varphi$ is a~bijection from $T_n$ of $(-m)$-loop case to $T^0_n$ of $(m+1)$-loop case. Combining it with the~construction described in \cite[Sec. 3.2]{KS1608}\footnote{In \cite{KS1608} LMOV invariants are considered instead of corresponding DT invariants, so $N_r(q)$ is used instead of $\Omega_r(q)$. The~definition of the~sets of modified Lyndon lists $T^{L,+}_r$ is given in  \cite[Eq. (3.32)]{KS1608})} we learn that 
\begin{equation}
\label{eq:PtoR}
\begin{split}
    \Omega_{\text{$(-m)$-loop}}(x',q) &= \sum_{r\geq 0}\left( \sum_{\phi\in (T^{L,+}_{r})_{\text{$(-m)$-loop}} }{\rm sgn}(\phi)q^{{\rm wt}(\phi)} \right)\frac{(x')^r}{[r]_{q^2}} \\
    &= \sum_{r\geq 0}\left( \sum_{\varphi\in (T^{L,+}_{r})_{\text{$(m+1)$-loop}} }{-\rm sgn}(\varphi)q^{2r - {\rm wt}(\varphi)} \right)\frac{(x')^r}{[r]_{q^2}} \\
    &= \sum_{r\geq 0}\left( \sum_{\varphi\in (T^{L,+}_{r})_{\text{$(m+1)$-loop}} }{-\rm sgn}(\varphi)q^{2 - {\rm wt}(\varphi)} \right)\frac{(x')^r}{[r]_{q^{-2}}} \\
    &= -q^2\Omega_{\text{$(m+1)$-loop}}(x',q^{-1})~,
\end{split}
\end{equation}
where we used
\begin{equation}
\begin{split}
    {\rm wt}(\varphi)&=\sum_{i=1}^n\varphi_i=2n-\sum_{i=1}^{n}\phi_i=2n-{\rm wt}(\phi)~,\\
    [r]_{q^2} &= \frac{1-q^{2r}}{1-q^2}~.
\end{split}
\end{equation}
In Ref.~\cite{KS1608} $\sum_{\varphi\in (T^{L,+}_{r})_{\text{$(m+1)$-loop}} }{\rm sgn}(\varphi)q^{- {\rm wt}(\varphi)}$ was linked to 
the $Q_r\left(q^{-2}\right)$ of Ref.~\cite{Rei11},
where $Q_r\left(q^{2}\right)$  was proven to be divisible by $[r]_{q^2}$ and that implies that $Q_r\left(q^{-2}\right)$ is also divisible by $[r]_{q^2}$. This happens because if $\zeta$ is a~root of unity then $\zeta^{-1}$ is also a~root of unity.

Now we only need to take into account that the~relation between $x'$ and $x$ for the~$(m+1)$-loop quiver with inverted $q$ reads $x=x'q^{m}$, therefore
\begin{equation}
    \Omega_{\text{$(-m)$-loop}}(x,q) = -q^{2}\,\Omega_{\text{$(m+1)$-loop}}(q^{-1}x,q^{-1})~,
\end{equation}
which is consistent with \eqref{eq:BPS_iso}.
\end{proof}

\bibliography{refs}
\bibliographystyle{JHEP}

\end{document}